\documentclass[Afour,sageh,times]{sagej}

\usepackage{moreverb,url}

\usepackage[colorlinks,bookmarksopen,bookmarksnumbered,citecolor=red,urlcolor=red]{hyperref}

\usepackage{lineno,hyperref}
\hypersetup{pdfauthor=author} %
\usepackage{booktabs}
\modulolinenumbers[5]
\usepackage{color}
\usepackage[dvipsnames]{xcolor}
\usepackage{comment}
\usepackage{verbatim}
\usepackage{float}
\usepackage{tikz}
\usetikzlibrary[shapes,shapes.arrows,arrows,shapes.misc,fit,positioning]
\usepackage[export]{adjustbox}
\usepackage{caption}
\usepackage{hyphenat}
\usepackage{subcaption}
\usepackage{caption}
\usepackage{siunitx}
\usepackage{amsmath}
\usepackage{makecell}
\usepackage{tabularray}
\usepackage{listings}
\usepackage{hyperref}
\usepackage[algo2e,vlined,nosemicolon]{algorithm2e}
\definecolor{codegreen}{rgb}{0,0.6,0}
\definecolor{codegray}{rgb}{0.5,0.5,0.5}
\definecolor{codepurple}{rgb}{0.58,0,0.82}
\definecolor{backcolour}{rgb}{.98,.98,.98}

\SetAlgoSkip{}
\SetAlCapSkip{0ex}

\makeatletter
\newcommand{\removelatexerror}{\let\@latex@error\@gobble}
\makeatother

\newcommand{\rev}[1]{{\color{black} #1}}

\begin{document}

\title{PETSc/TAO Developments for \rev{GPU-Based} Early Exascale Systems}

\author{Richard Tran Mills\affilnum{1},
Mark Adams\affilnum{2},
Satish Balay\affilnum{1},
Jed Brown\affilnum{3},
Jacob Faibussowitsch\affilnum{1}, %
Toby Isaac\affilnum{1},
Matthew Knepley\affilnum{4},
Todd Munson\affilnum{1},
Hansol Suh\affilnum{1},
Stefano Zampini\affilnum{5},
Hong Zhang\affilnum{1},
and Junchao Zhang\affilnum{1}
}
\affiliation{\affilnum{1}Argonne National Laboratory, Lemont, Illinois, USA\\
\affilnum{2}Lawrence Berkeley National Laboratory, Berkeley, California, USA\\
\affilnum{3}University of Colorado Boulder, Colorado, USA\\
\affilnum{4}University at Buffalo, Buffalo, New York, USA\\
\affilnum{5}King Abdullah University of Science and Technology, Thuwal, Saudi Arabia\\
}

\corrauth{Richard Tran Mills, Mathematics and Computer Science Division, Argonne National Laboratory, Lemont, Illinois, USA.}
\email{rtmills@anl.gov}

\begin{abstract}
The Portable Extensible Toolkit for Scientific Computation (PETSc) library provides scalable solvers for nonlinear time-dependent differential and algebraic equations and for numerical optimization via the Toolkit for Advanced Optimization (TAO). PETSc is used in dozens of scientific fields and is an important building block for many simulation codes.
During the U.S. Department of Energy's Exascale Computing Project, the PETSc team has made substantial efforts to enable efficient utilization of the massive fine-grain parallelism present within exascale compute nodes and to enable performance portability across exascale architectures.
We recap some of the challenges that designers of numerical libraries face in such an endeavor, and then
discuss the many developments we have made, which include the addition of new GPU backends, features supporting efficient on-device matrix assembly, better support for asynchronicity and GPU kernel concurrency, and new communication infrastructure.
We evaluate the performance of these developments on some pre-exascale systems as well the early exascale systems {\it Frontier} and {\it Aurora}, using compute kernel, communication layer, solver, and mini-application benchmark studies,
and then close with a few observations drawn from our experiences on the tension between portable performance and other goals of numerical libraries.

\end{abstract}


\maketitle

\section{Introduction} %
In little more than a decade, we have seen a shift in high-performance computing (HPC)
from complete reliance on central processing units (CPUs) to the incorporation of graphics processing units (GPUs)
that provide the bulk of the computing power in most supercomputers.
This kind of heterogeneous architecture with CPUs and GPUs
is exemplified by the several high-end pre- or early exascale machines
funded by the U.S. Department of Energy (DOE),
including Summit at the Oak Ridge Leadership Computing Facility (OLCF) and Perlmutter at
the National Energy Research Scientific Computing Center (NERSC) with NVIDIA GPUs,
Frontier at OLCF and El Capitan at Lawrence Livermore National Laboratory (LLNL) with AMD GPUs, and
Aurora at the Argonne Leadership Computing Facility (ALCF) with Intel GPUs.
Exascale machines, with the capability to perform $10^{18}$ floating point operations per second, represent a 1000x leap from previous petascale machines.
The DOE Exascale Computing Project (ECP)~\citep{ecp-kothe-lee-qualters-2019} aimed to prepare DOE mission-critical applications, an integrated software stack,
and exascale hardware technology to form a capable exascale computing ecosystem.
Although GPUs, sometimes referred to as ``accelerators'' or  ``devices'' (in contrast to CPU ``hosts''), have higher throughput and energy-efficiency
than CPUs, they also bring profound programming challenges due to their architectural departures from CPUs and still-evolving software environments.
GPUs use massive fine-grain parallelism, and thus programmers must write massively parallel code at the intra-node level in order to effectively utilize the compute power.
GPUs usually have discrete device memory in addition to the main memory used by CPUs, and often the device and host cannot directly dereference pointers for the other memory.
Programmers have to manage these two kinds of memory and differentiate pointers between them, e.g., when passing pointer arguments to routines.
Though unified shared memory (USM), available on modern heterogeneous systems, enables
the creation of a single address space accessible to both the CPU and GPU,
simplifying memory management and data sharing between these processors,
it may not provide transparent performance:
programmers often must prefetch data to avoid performance penalties.
More importantly,
computational kernels on device are executed asynchronously with respect to the host process on CPUs, and
kernels can be launched to different streams (like work queues) on a GPU and executed independently.
Programmers must be aware of the asynchronicity and GPU streams, and synchronize their computations on host or on device properly to
maintain correct code.

To program GPUs, developers can use vendor-provided programming models and libraries,
such as CUDA, cuBLAS, and cuSparse from NVIDIA \citep{CUDA};
HIP, hipBLAS, and hipSparse from AMD \citep{HIP};
and SYCL and OneAPI from Intel \citep{SYCL}.
CUDA is proprietary and currently the most popular programming model for general-purpose processing on accelerators thanks to NVIDIA's large market share.
HIP, syntactically very similar to CUDA, allows developers to create portable applications for AMD and NVIDIA GPUs from a single source code.
On AMD GPUs, hipBLAS and hipSparse are simply wrappers over the rocBLAS and rocSparse libraries underneath the AMD ROCm software stack for GPU programming.
SYCL is a stark departure from CUDA and HIP and is advertised as
an open standard allowing developers to use a C++ single source to program a wide range of targets,
including CPUs, GPUs, digital signal processors (DSPs), and field programmable gate arrays (FPGAs).
AMD and Intel also provide translation tools, such as \textit{hipify} from AMD and \textit{SYCLomatic} from Intel, to
automatically convert source codes from CUDA to HIP or SYCL, respectively.
Though these migration tools are helpful, none can guarantee automatic translation.
Usually, programmers have to intervene to fix untranslatable code.
To enhance performance portability across various GPUs and CPUs,
ECP  funded several projects such as RAJA \citep{RAJA}, Kokkos \citep{KOKKOS}
and OpenMP target offload \citep{OpenMP} compilers.
RAJA and Kokkos use C++ templates and lambda functions to provide a portability layer in the front end
but rely on vendor programming models and compilers in the backend.
OpenMP target offload uses preprocessor directives (pragmas) in source code and thus requires compiler support.
OpenMP has been a popular choice for Fortran applications, as Fortran support is poor in the other models.
With all these options available, ECP applications and libraries
were not required to use a certain programming model.
The variety of programming languages, code complexity, varying package dependencies, and large amount of legacy
code precludes a one-size-fits-all solution.

PETSc/TAO (PETSc for short) \citep{petsc-3.21}, one of the math libraries funded by ECP,
features scalable solvers for nonlinear time-dependent differential
and algebraic equations and for numerical optimization.
PETSc is mainly written in the C programming language, though employing an object-oriented design.
PETSc provides Fortran and Python bindings and interfaces with dozens of external packages in various languages,
such as the direct solvers MUMPS (in Fortran) \citep{mumps01} and SuperLU (in C) \citep{superlu}, the
multigrid solver Hypre (in C) \citep{hypre-users-manual},
and the adaptive mesh refinement framework AMReX (in C++) \citep{AMReX_JOSS}.
PETSc, which has formed an ecosystem with applications, libraries, and frameworks built upon it~\citep{petsc-community2022}, is widely used in academia, government laboratories, and industry and
is an important building block for many simulation codes. Likewise, PETSc is part of the xSDK~\citep{xsdk:homepage} and E4S~\citep{e4s:homepage} ecosystems of complementary HPC packages.

Because PETSc functions as numerical infrastructure for many application packages,
the library design completely separates the programming
models used by applications (or external packages) and the models used by PETSc
for its backend computational kernels.
This flexibility is accomplished by sharing data between the application and PETSc programming models but not
sharing the programming models' internal data structures.
Because the data is shared, there are no copies between models and no loss of efficiency.
This separation allows PETSc users from C/C++, Fortran, or Python to employ their preferred
GPU programming model, whether it be CUDA, HIP, or Kokkos.
In all cases, users can rely on PETSc's large assortment of composable, hierarchical, and nested
solvers \citep{bkmms2012}, as well as advanced time-stepping and adjoint
capabilities \citep{Zhang2022tsadjoint} and numerical optimization methods running on the GPU. For example, an application for solving time-dependent partial differential
equations may compute the Jacobian using Kokkos
and then call PETSc's time-stepping
routines and algebraic solvers that use CUDA.
Applications can mix and match programming models,
allowing, for example, some application code in HIP and some in CUDA.

In \cite{mills2021toward}, we provided a blueprint for porting PETSc
applications to use GPUs, surveyed challenges in developing efficient and portable mathematical
libraries for GPU systems, and introduced the PETSc backend developments meant to meet
these challenges at that time.
With the sunset of ECP and the availability of early exascale computers such as Frontier and Aurora,
we summarize the accomplishments of the PETSc/TAO ECP project,
and we present the latest developments on performance portability through PETSc,
illustrated by results on exascale computers.
This paper is organized as follows.
We first recap some GPU programming challenges and PETSc's responses in Section \ref{sec:challenges}, then
provide an update in Section \ref{sec:petsc_port} to the PETSc application GPU-porting blueprint.
Afterwards, we introduce in detail
\rev{
some lower-level PETSc developments for GPUs, including improved communication support (including asynchronous,
GPU-initiated communication) and efficient matrix assembly in Sections \ref{sec:petscsf} and \ref{sec:matassembly}.
Next, we move higher in the PETSc software hierarchy and discuss two approaches that can increase GPU throughput and
utilization for linear solves: asynchronous solvers and batched solvers (Sections \ref{sec:asynchronous} and
\ref{sec:batched_solvers}, respectively).
We then discuss the use of GPU-friendly dense reformulations of algorithms to improve the performance of
the limited memory BFGS method (L-BFGS) in Section \ref{sec:dense-formulate}.
The paper culminates in two mini-application studies using PETSc on GPUs: solving the 3D Laplacian using algebraic
multigrid (Section \ref{sec:gamg}) and time evolution of the Landau collision integral (Section \ref{sec:landau}).
We present some brief conclusions and thoughts towards the future in Section \ref{sec:conclusion}.
}
We note that, throughout the paper, we use the term \textit{programming model} to refer to
both the model and its supporting runtime.
With one exception, our experiments were conducted on the four (pre-) exascale machines
with configurations shown in Table \ref{tab:machines}.

\begin{table*}
    \centering
    \small
        \caption{Configurations of the (pre-)exascale machines we used for experiments.
        $^a$Note that we treat the two graphics compute dies (GCDs) in an AMD MI250X on Frontier or the two tiles in an Intel PVC on Aurora as two separate GPUs.
        $^b$We report unidirectional bandwidth numbers.
        $^c$Measured by a benchmark since we are unaware of the theoretical bandwidth.
        }
    \begin{tblr}{width=\textwidth,|Q[m]|Q[m]|Q[m]|Q[m]|Q[m]|}
    \hline
        Machine & Summit@OLCF &  Perlmutter@NERSC &  Frontier@OLCF &  Aurora@ALCF\\ \hline
        CPUs per node&  2x IBM Power9 &  1x AMD EPYC-7763 & 1x AMD EPYC-7763  & 2x Intel Xeon SPR \\
        GPUs per node& 6x  NVIDIA Tesla V100 &  4x NVIDIA A100 & {4x AMD MI250X \\(2 GCDs per GPU)} & {6x Intel GPU PVC\\(2 tiles per PVC)} \\
        HBM per GPU$^a$ &  16 GB, 900 GB/s &  40 GB, 1.5 TB/s&  64 GB, 1.6 TB/s & 64 GB, 1.6 TB/s\\
         {GPU-GPU link$^b$}  &  NVLINK, 50 GB/s &  NVLINK-3, 100 GB/s&  Infinity Fabric, 50{$\sim$}200 GB/s & Xe Link, 15{$\sim$}196 GB/s$^c$\\
         Network & InfiniBand, Fat-tree & Slingshot-11, Dragonfly & Slingshot-11, Dragonfly & Slingshot-11, Dragonfly \\
        Software &  {IBM Spectrum MPI-10.3,\\ cuda-10.1, gcc-6.4\\NVSHMEM-2.1.2}&  {Cray MPICH-8.1.28,\\ cuda-12.2, gcc-12.3, \\ Kokkos-4.3}
        & {Cray MPICH-8.1.23,\\ rocm-5.4, gcc-12.2, \\ Kokkos-4.3} & {Aurora MPICH-52.2, \\ oneapi-2023.12.15.001,\\ Kokkos-4.3} \\ \hline
    \end{tblr}
    \label{tab:machines}
\end{table*}

\section{GPU programming challenges and PETSc's responses} \label{sec:challenges}
In \cite{mills2021toward}, we enumerated three fundamental GPU programming challenges:
\begin{enumerate}
    \item [F1] {\it Portability of application codes:} write code that is portable across different hardware.
    \item [F2] {\it Algorithms for high-throughput systems:} design parallel algorithms exploiting high GPU-concurrency.
    \item [F3] {\it Utilizing all GPU and CPU compute power:} reduce idle and waiting time to improve hardware utilization.
\end{enumerate}
In addition, we discussed seven ancillary challenges:
\begin{enumerate}
    \item [A1] {\it Managing the kernel queue:} pipeline many, not-too-small kernels in GPU streams to keep GPUs busy.
    \item [A2] {\it Network communication:} communication should be stream-aware without the need to synchronize GPUs.
    \item [A3] {\it Over- and undersubscription:} overcome the mismatch between the number of CPU cores and GPUs.
    \item [A4] {\it CPU-GPU communication time:} lower cost by reducing the communication amount or overlapping communication with computation.
    \item [A5] {\it Multiple memory types:} manage different types of memory either explicitly or implicitly via USM.
    \item [A6] {\it Use of multiple streams from libraries:} manage streams within a library or across libraries while maintaining data dependence.
    \item [A7] {\it Multiprecision on the GPU:} take advantage of the higher compute power of GPUs with lower precision.
\end{enumerate}

While F1 concerns {\it portability}, A5 and A6 are about {\it correctness},
and F2, F3, A1--A4 and A7 are related to {\it performance}.
To meet these challenges, during ECP we greatly improved
PETSc's GPU capability around these three aspects
with new GPU backends and constructs, new GPU-friendly application programming interfaces (APIs),
and new algorithms for high-throughput computation.
One can refer to \cite{mills2021toward} for categorized PETSc responses to these challenges.
Here, we further explain the issues of portability, correctness, and multiprecision with respect to the latest developments in PETSc.

\paragraph{On PETSc's code portability:}
Before ECP, we had a legacy PETSc/CUDA backend for NVIDIA GPUs,
and a PETSc/OpenCL backend via ViennaCL \citep{VIENNACL} for OpenCL devices.
We implemented vector and matrix sub-types for these backends, such as {\tt VECCUDA} and {\tt MATAIJVIENNACL},
and encouraged users to use PETSc's options database from the command line, e.g., {\tt -vec\_type cuda} or
{\tt -mat\_type aijviennacl} to set GPU-specific object types at runtime. Using these options, the same PETSc application source code
can work with either a CUDA device or an OpenCL device.
We did not write many device kernels ourselves.  Instead, we relied on vendor libraries
to provide basic vector and matrix operations at the process level. PETSc mainly managed the MPI parallelism and host-device synchronization.
During ECP, we also needed to support AMD and Intel GPUs.
Ideally, we should use a unified, portable programming model for all devices,
but unfortunately, no suitable model existed then.
OpenCL has portable performance as an overarching goal, 
\rev{
but the API is verbose, and vendor support for OpenCL in compilers and performance tools is limited \citep{colleen2020}.
}
Therefore, for NVIDIA GPUs, we further optimized and expanded the PETSc/CUDA backend with new features.
Later, staff from AMD helped us add a PETSc/HIP backend for AMD GPUs, which was very similar to the PETSc/CUDA backend thanks to
the similarity between HIP and CUDA.
We partially consolidated the PETSc/CUDA and the PETSc/HIP backends via a C++ abstraction layer we developed named CUPM (\textbf{CU}DA-like \textbf{P}rograming \textbf{M}odel)\rev{, which hides CUDA/HIP API disparities in PETSc. }
Further work is needed to fully consolidate the two backends.
We did not use hipify, the AMD CUDA to HIP migration tool,
because PETSc supports multiple versions of CUDA and contains many CUDA library calls to cuBLAS and cuSparse, which makes
the translation more difficult.
In addition to these vendor-native models, we developed a new GPU backend using the Kokkos programming model and
employing Kokkos-Kernels (a math library providing BLAS, sparse BLAS etc.) for basic math operations.
We currently do not have a RAJA implementation, as it lacks a math kernel library similar to Kokkos-Kernels.
The PETSc/Kokkos backend can run on various CPUs and GPUs, including NVIDIA, AMD, and Intel.

Going forward, as portable device programming models mature, we hope to consolidate these PETSc GPU backends to reduce maintenance burden and avoid code duplication.
However, internal reorganization will not affect existing PETSc users working with PETSc objects.
As mentioned before,
we share data between an application and the PETSc programming model, but do not share the programming
models’ internal data structures. We do provide utility APIs for PETSc users employing some popular GPU programming models.
For example, for Kokkos users, we have APIs returning a {\tt Kokkos::View} from a PETSc vector; for CUDA users,
given the same PETSc vector, we have APIs returning a device pointer pointing to the vector storage in device memory.

\paragraph{On PETSc's memory types:}
PETSc assumes discrete device memory and maintains two copies of data, one on the host and the other on the device,
for PETSc GPU vectors ({\tt Vec}) and matrices ({\tt Mat}). These objects have an internal mask indicating where the latest data resides.
If a given Vec or Mat operation is not implemented on the device, we implicitly synchronize the copy on the host and perform the operation there instead, so that we always have full API compliance.
\rev{
Unintended implicit memory copies are a potential performance pitfall,
but PETSc's built-in lightweight performance profiling scheme can assist in identifying such issues:
Users can specify the {\tt -log\_view} runtime option and search the log for unexpected high counts or large sizes of copies between host and device per PETSc event.
}
PETSc {\tt Vec/Mat} APIs identify themselves as either read-only, write-only, or read-and-write on parameters. With this information, we can update the
mask accordingly and avoid unnecessary memory copying between the host and device.
For users who want to directly obtain the device array used by PETSc objects, we provide APIs such as
{\tt VecGetArrayAndMemType\{Read,Write\}(Vec x, PetscScalar **a, PetscMemType *mtype)},
with the returned pointer {\tt a} pointing to the vector array and
{\tt mtype} being one of {\tt PETSC\_MEMTYPE\_\{HOST, CUDA, HIP, SYCL\}}. For PETSc GPU objects, calling these routines always returns the latest data on the GPU,
while for CPU objects, it returns the array on the host.
Callers can retrieve {\tt a}'s memory type from the {\tt mtype} argument.
Similarly, we also provide APIs for creating PETSc device objects with user-provided device arrays.
This approach will work even when CPUs and GPUs share the same physical
memory, as in the upcoming El Capitan supercomputer at LLNL.
In this case, we will have a single copy of data, making host-device memory copying a no-op.

\paragraph{On PETSc's GPU streams:}
Streams are an important mechanism on GPUs to hide serial kernel launch latency by pipelining kernel launches.
Streams can also improve resource utilization with extra parallelism by potentially executing
kernels in different streams concurrently.
In the latter sense, GPU streams are like CPU threads, but the programming paradigm is very different.
Suppose that there are two functions on the host, a caller and a callee.
With CPU threads, the callee is automatically executed on the same thread as the caller.
With GPU streams, the caller must pass the callee the stream that it uses by some mechanism, e.g.,
via an argument, through a global variable, or they may both assume a default stream.
For an existing library written in C++, one might have the luxury of overloading all
its API functions with an extra stream argument without breaking existing code.
Though doable, it is not only daunting but also error-prone because users might
call an original API when they are supposed to call the one with a stream argument,
resulting in mismatched streams being used.
For a C library like PETSc, we cannot afford to break the API or
bloat it with a host of new functions merely by adding an additional stream argument.
Thus, at the very basic level,
we use a variable storing a PETSc global stream and wrap it into a generic object of type
{\tt PetscDeviceContext}.
{\tt PetscDeviceContext} provides \textit{uniform APIs} across different PETSc GPU backends such as CUDA, HIP and SYCL.
Users can query the device type to get a handle to the stream, and then cast the handle to the specific stream type according to the
device type, as shown in the code below. For Kokkos users,
we provide a utility function {\tt PetscGetKokkosExecutionSpace()} to directly return
a Kokkos execution space instance with the stream.
\begin{lstlisting}[label={lst:PetscDeviceContext},]
  PetscDeviceContext dctx;
  PetscDeviceType    type;
  void               *handle;

  PetscDeviceContextGetCurrentContext(&dctx);
  PetscDeviceContextGetDeviceType(dctx, &type);
  PetscDeviceContextGetStreamHandle(dctx, &handle);

  if (type == PETSC_DEVICE_CUDA) {
    cudaStream_t stream = *(cudaStream_t *)handle;
    myKernel<<<64, 128, 0, stream>>>();
  } else if (type == PETSC_DEVICE_HIP) {
    hipStream_t stream = *(hipStream_t *)handle;
    ...
  }
\end{lstlisting}
From the user's point of view, PETSc works as if using one global GPU stream.
\rev{
(If we find that users wish to make PETSc calls from multi-threaded GPU user code, we could add an option to
protect this global state with locks or to store it in thread-local storage, analogously to
how an OpenMP thread-safe subset of PETSc can be configured with the {\tt --with-threadsafety} option).
}
PETSc users should query this stream with the above methods when they want to
coordinate with PETSc streams.
The type of stream can be changed globally with the command line option {\tt -root\_device\_context\_stream\_type}.
We are exploring multi-stream parallelism with 
PETSc device contexts, see Section \ref{sec:PetscDeviceContext}.

\rev{
\paragraph{On PETSc's multiprecision support:}
Partly because of PETSc's origins as a C library, it currently supports only one precision that is set at configuration time.
This precision---which could be single, double, float128, or fp16---defines {\tt PetscScalar}, the data type of \textit{all} PETSc vector and matrix entries. 
Enforcing a uniform data type simplifies the code design and helps to prevent bugs and improve performance.
Some users, however, need to call certain external packages (e.g., the direct solver SuperLU \citep{superlu}) in lower precision to save memory.
Furthermore, the proliferation of GPUs supporting higher performance in lower precisions makes it desirable to enable some multiprecision support in PETSc.
For a few external packages such as SuperLU and HPDDM \citep{jolivetromanzampini2020}, we have already provided options to use a precision other than {\tt PetscScalar}. 
For example, one can use {\tt -pc\_factor\_mat\_solver\_type superlu\_dist -pc\_precision single} to call SuperLU in single while PETSc is in double precision.
PETSc will do the precision conversion automatically when building data structures for the packages. 
We can adopt the same approach when offloading data to GPUs.
As mentioned above, PETSc maintains two copies of data: one on the host and the other on the device.
Currently, they are in the same precision. 
We plan to provide new APIs to let users specify a different precision for data on device for PETSc GPU {\tt Vec/Mat} types, either globally or at the linear solver level.
This will allow some multiprecision support for GPUs, without greatly increasing code complexity and without affecting the CPU user API.
}

\section{Porting PETSc applications to GPUs} \label{sec:petsc_port} %

\cite{mills2021toward} introduced general strategies for porting PETSc applications to GPUs and recommended an
incremental approach facilitating easy comparison of GPU and CPU results.
Here, we summarize the approach we presented, with some updates reflecting the latest PETSc performance portability design.
We mainly focus on GPU work and use an application using Kokkos as an example.

Listing \ref{lst:main} displays an excerpt of a typical PETSc main application program
for solving a nonlinear set of equations on a structured grid using Newton's method.
This example illustrates the common usage pattern that applies when using components
of PETSc, whether time integrators, nonlinear solvers, linear solvers, etc.:
\begin{itemize}
    \item Setup application data, meshes, initial state, etc., (here, a {\tt SNES} solver object, a data management object {\tt DM} for a 1-D domain,
          a vector of degrees of freedom {\tt Vec}, and a {\tt Mat} to hold the Jacobian),
    \item provide a callback for the {\tt  Function} that defines the problem (the nonlinear residual),
    \item provide a callback for the {\tt  Jacobian} of the {\tt Function},
    \item call the PETSc solver ({\tt SNESSolve()}), possibly in a loop.
\end{itemize}
This pattern holds whether targeting CPUs or GPUs for execution.
When porting to GPUs, the creation and manipulation of solver, matrix, and vector objects do not change,
but users will need to 1) write code to ensure that needed data structures are either copied from CPU memory to device memory
or constructed directly on the GPU and 2) provide {\tt Function} and (optionally) {\tt Jacobian} routines that call GPU kernels.
We recommend adopting an incremental approach, in which portions of the computation are moved to GPU and then evaluated
for correctness and performance; porting can be considered complete when observed GPU speedup is sufficient, relative to the
cost of the rest of the simulation run.

\begin{lstlisting}[caption={Main application code, with no syntactic changes from typical CPU version.},label={lst:main},captionpos=b]
SNESCreate(PETSC_COMM_WORLD,&snes);
DMDACreate1d(PETSC_COMM_WORLD,...,&ctx.da);
DMCreateGlobalVector(ctx.da,&x);
VecDuplicate(x,&r);
DMCreateMatrix(ctx.da,&J);
SNESSetFunction(snes,r,KokkosFunction,&ctx);
SNESSetJacobian(snes,J,J,KokkosJacobian,&ctx);
SNESSolve(snes,NULL,x);
\end{lstlisting}

Listing \ref{lst:function} shows a Kokkos implementation of {\tt Function}, which is similar to the traditional CPU version.
For simplicity, we assume periodic boundary conditions in one dimension, but the pattern is similar in more dimensions and with general boundary conditions.
{\tt DMDAVecGetKokkosOffsetView(da, xl, \&X)} returns
a Kokkos OffsetView {\tt X} from a PETSc vector {\tt xl}.
Because {\tt X} and {\tt xl} share data,
we wrap but do not copy {\tt xl}'s data.
Kokkos OffsetViews are essentially multi-dimensional arrays
with non-zero start indices. PETSc uses them so that users can access a locally owned
array with conceptually simpler global indices. Via the mesh management object {\tt da},
PETSc knows the dimension, start, and end indices of the OffsetView so that
the returned object is properly populated.
Then, users can write parallel device code of their choice with these OffsetViews.
A critical issue here is that users must be aware of the
asynchronous nature of GPU computations, and therefore must know the current GPU stream used by PETSc.
PETSc Kokkos users have the {\tt PetscGetKokkosExecutionSpace()} to get a Kokkos execution space instance that wraps the stream.
With that, one can construct a Kokkos {\tt RangePolicy} object, for example in
a Kokkos {\tt parallel\_for} dispatch,
so that the user's device code and PETSc's share the same stream, and to enforce data dependency.
\begin{lstlisting}[caption={{
 {\tt Function} callback code.
\tt xl}, {\tt x}, {\tt r}, {\tt f} are PETSc {\tt Vec}s, while {\tt X}, {\tt R},
 {\tt F} are Kokkos {\tt OffsetView}s. Marking {\tt R} write-only offers PETSc an optimization hint.
 Note that via {\tt exec} user shares the GPU stream that PETSc is using.
 },
 label={lst:function},captionpos=b]
DMGetLocalVector(da,&xl);
DMGlobalToLocal(da,x,INSERT_VALUES,xl);
DMDAVecGetKokkosOffsetView(da,xl,&X); // no copy
DMDAVecGetKokkosOffsetViewWrite(da,r,&R); // R is W-only
DMDAVecGetKokkosOffsetView(da,f,&F);
PetscInt xs = R.begin(0), xm = R.end(0);
auto exec = PetscGetKokkosExecutionSpace();
Kokkos::parallel_for(
  Kokkos::RangePolicy<>(exec,xs,xm),[=](PetscInt i){
    R(i) = d*(X(i-1)-2*X(i)+X(i+1))+X(i)*X(i)-F(i);});
\end{lstlisting}

The Jacobian computation departs from the block-oriented {\tt MatSetValues()} approach traditionally used in CPU-based
PETSc codes, as it does not map efficiently onto GPU architectures.
We discuss this topic in detail in Section \ref{sec:matassembly}, and here simply note that to use
the new coordinate-based {\tt MatSetValuesCOO()} approach each GPU thread
places the non-zeros it contributes into a device buffer and later calls
{\tt MatSetValuesCOO()} to have PETSc post-process the buffer, as shown in Listing \ref{lst:jacobian}.
\begin{lstlisting}[caption={{\tt Jacobian} callback code. {\tt v}, the Kokkos View managed by users,
works as a device buffer to store unprocessed nonzeros calculated by GPU threads.
},
label={lst:jacobian},captionpos=b]
DMDAVecGetKokkosOffsetView(da,x,&X);
PetscInt xs = R.begin(0), xm = R.end(0);
auto exec = PetscGetKokkosExecutionSpace();
auto v = ctx.v; // a Kokkos::View used as a device buffer
Kokkos::parallel_for(
  Kokkos::RangePolicy<>(exec,xs,xm),[=](PetscInt i){
    PetscInt ofst = (i-xs)*3; // offset in v for thread i
    v(ofst+0) = d;
    v(ofst+1) = -2*d + 2*X(i);
    v(ofst+2) = d;});
MatSetValuesCOO(J,v.data());
\end{lstlisting}

\section{Communication on GPUs} \label{sec:petscsf} %
As PETSc's strength lies in distributed computation with sparse and irregular data, the library has to take care of complex communication among
processes, such as those in sparse matrix-vector multiplication, sparse matrix-matrix multiplication, or irregular mesh partitioning.
In PETSc, we designed a module named \textit{PetscSF} \citep{PetscSF_TPDS_2021} to encapsulate
frequently used communication operations.
Underneath a unified interface, PetscSF can have different communication implementations, though
the default uses MPI.
In this section, we first introduce
PetscSF's design and the extensions to allow MPI communication with device data.
Next, we introduce an experimental PetscSF implementation using NVIDIA NVSHMEM \citep{NVSHMEM} that can overcome some limitations imposed by MPI.
Finally, we use a microbenchmark written with PetscSF to study
 \rev{the GPU-aware MPI} communication performance on the four target machines.

\subsection{The star-forest (SF) abstraction}
\begin{figure}[htbp]
\begin{center}
\includegraphics[width=1.0\linewidth]{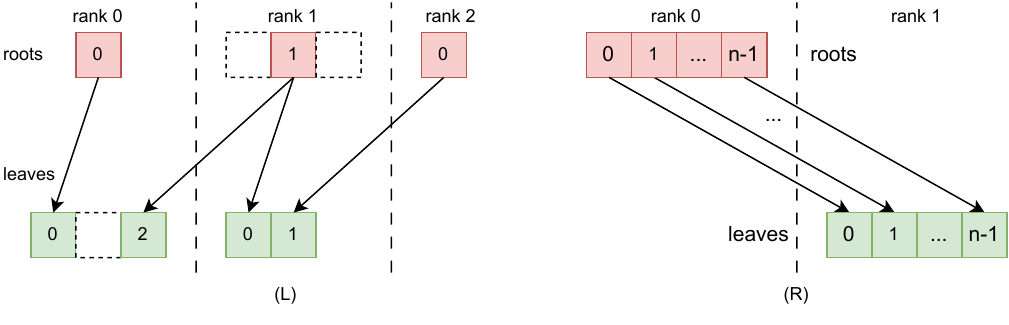}
\caption{Two star-forest examples. The left example has three MPI ranks, while the right has two.
Vertical dashed lines separate MPI ranks.
Colored boxes are roots (leaves). Enclosed numbers are indices of the roots (leaves) in their index space.
Dashed boxes represent holes in the spaces not belonging to the SF.}
\label{fig:SF}
\end{center}
\end{figure}

PetscSF uses {\it star-forests} to abstract communication patterns.
A star is a simple tree with one root vertex connected to zero or more leaves.
A star forest is a disjoint union of stars; see examples in Figure \ref{fig:SF}.
Leaves are locally indexed with integers, while roots are globally indexed via tuples of (owner rank, offset).
 A \texttt{PetscSF} is created collectively by specifying, for each leaf on the current process,
 the owner rank and an offset of the corresponding root on the owner.
 PETSc analyzes the graph and derives the communication pattern.
We provide APIs to communicate between roots and leaves, for example:
\begin{lstlisting}
PetscSFBcastBegin/End(PetscSF sf, MPI_Datatype unit,
  const void *rootdata, void *leafdata, MPI_Op op);
PetscSFReduceBegin/End(PetscSF sf, MPI_Datatype unit,
  const void *leafdata, void *rootdata, MPI_Op op);
\end{lstlisting}

The former broadcasts root values to leaves, while the latter reduces leaf values into roots, with
both taking a reduction operation specified by an
\texttt{MPI\_Op} argument to \textit{add} the source values or to replace the destination values (with \texttt{op=MPI\_REPLACE}).
The \texttt{Begin/End} split-phase design allows users to
insert computations in between to potentially overlap computation with communication.
It is common in irregular applications that leaves and roots are not consecutive in their respective index space.
In that case, PETSc will call its pack or unpack kernels to put data into internally managed buffers for send and receive.
The same PetscSF can be applied to different root/leaf data or data types.
Depending on the communication pattern, PetscSF can perform the operation using MPI point-to-point or collective operations.
For the most common sparse neighborhood communication pattern, we use persistent MPI sends and receives by default,
but we also support MPI neighborhood collectives.

\subsection{The mismatch between MPI and GPUs} \label{subsec:mpi-gpu-mismatch}
\begin{figure}[htbp]
\begin{center}
\includegraphics[width=.9\linewidth]{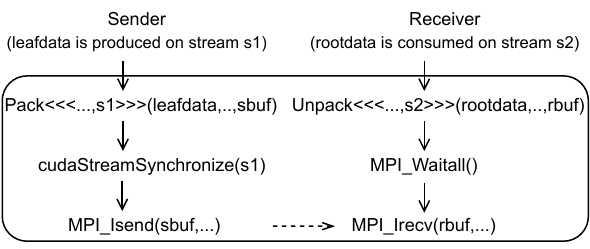}
\caption{A typical data path of {\tt PetscSFReduce()} with CUDA, assuming
all parts except MPI work on a common device stream local to the calling process.
Note the stream synchronization before {\tt MPI\_Isend()}.}
\label{fig:SF-SyncModel}
\end{center}
\end{figure}

As more and more computations are offloaded to the device for acceleration,
it is desirable to directly communicate data on the device instead of staging data on the host merely for the purpose of communication,
because moving data between host and device is expensive.
GPU-aware MPI was introduced to solve this problem by allowing MPI calls to
accept device buffers.
However, one has to be aware that GPU computations or kernels are queued in streams and are executed
asynchronously with respect to the host process.
MPI routines, as a host side API, must make sure that the send data on the device, which could be the output of some previous GPU kernels,
is ready to be sent.
We would like to inform MPI of the stream in use so that it can maintain the data dependency.
Unfortunately, as of this writing, the MPI standard does not support this functionality since MPI routines do not take a GPU stream argument.
As a result, MPI users must synchronize the GPU stream to have the send data ready:
In other words, force all kernels producing the send data to complete before launching new kernels.
On the receiver side, an MPI receive must block subsequent GPU kernel launches, i.e., synchronize the device.
In PETSc, we have a default device stream. With that, Figure \ref{fig:SF-SyncModel} gives a
typical data path when using CUDA for \texttt{PetscSFReduce()}, which reduces sparse leaf data to root data with the help of the Pack/Unpack kernels.

Because kernel launches are expensive, the GPU runtime provides users the stream mechanism for
pipelining kernel launches, so that kernel launch latency can be hidden by the execution
of previous kernels. MPI incurs a device synchronization that stalls the pipeline, making
the cost of kernel launches harder to mitigate.
To quantify the launch latency, we designed a microbenchmark, in which we launch an empty GPU kernel many times,
marking on the CPU before and after each launch, and then calculating the average latency per launch (i.e., per iteration).
We had two variants: in one (Asynchronous), there were no stream synchronizations at all;
in the other (Synchronous), we inserted a stream synchronization after each launch.
The results are shown in Table \ref{tab:launch},
from which we can see the huge impact of synchronization on kernel launch latency.
\begin{table}[ht]
\small
\caption{Average latency ($\mu s$) of launching an empty kernel asynchronously or synchronously on (pre-)exascale machines}
\begin{tabular}{|l|r|r|r|r|}
\hline
Platform      & Summit   & Perlmutter   & Frontier   & Aurora   \\ \hline
Asynchronous  & 4.9 & 2.3  & 1.9   & 3.3    \\ \hline
Synchronous   & 12.8 & 7.1  & 7.8   & 6.2    \\ \hline
\end{tabular}
\label{tab:launch}
\end{table}

\subsection{Synchronization-free PetscSF with NVSHMEM} \label{petscsf-nvshmem}
To avoid the device synchronizations imposed by MPI,
we developed an experimental implementation of PetscSF using NVSHMEM \citep{NVSHMEM}.
NVSHMEM is NVIDIA's implementation of OpenSHMEM \citep{OpenSHMEM} on CUDA devices.
The OpenSHMEM standard specifies an API for partitioned global address space (PGAS) parallel programming,
providing one-sided shared-memory style \textit{put}/\textit{get} APIs to access remote objects.
NVSHMEM supports point-to-point and collective communications between GPUs within a node or
over networks.
Communication can be initiated either on the host or on the device.
Unlike MPI, NVSHMEM host-side APIs take a stream argument.
Remotely accessible objects (aka symmetric objects) are collectively allocated
over all processing elements (PEs, like processes in MPI) from a special heap (called the symmetric heap).
All PEs must allocate a symmetric object with the same size, so that the object always
appears at the same offset in their symmetric heap.
PEs access remote data by referencing a symmetric address and the rank of the \textit{remote} PE.
A symmetric address is the address of a symmetric object on the \textit{local} PE, plus an offset if needed.
In the NVSHMEM implementation of PetscSF, we still pack/unpack data in send/receive buffers, but those buffers
are now allocated as symmetric objects on the device. We launch, on the PETSc default stream, a CUDA kernel which then calls the device function \texttt{nvshmem\_putmem\_nbi} to put data from send buffers into receive buffers.
NVSHMEM can be used with MPI. \rev{In our design, we use MPI to exchange information, such as sizes and offsets of buffers,
required to set up the communication, and then we use NVSHMEM to carry it out.}
We found implementing PetscSF in NVSHMEM no less complex than in MPI.
Symmetric allocation forced us to take the maximal size of send/receive buffers across all PEs, though their real sizes could vary.
Shared-memory style APIs forced us to design complex communication protocols
even for simple patterns, such as knowing when data in a send buffer
has already been sent and thus is safe to overwrite, or when data in a receive buffer
is ready for use. The detailed design was published in \cite{PetscSF_TPDS_2021}.
We hope that the MPI Forum will soon provide GPU-initiated communication support in the
MPI standard, such as in the proposal of \cite{zhou2022mpix}, which would greatly simplify implementation.
Currently, our PetscSF implementation can do stream-aware communication across CUDA devices, free of device synchronization.
In Section \ref{subsec:cg_nvshmem}, we have used it to implement a parallel asynchronous conjugate gradient (CG) solver.

\subsection{\rev{GPU-aware MPI} message passing latency on (pre-)exascale machines}
\begin{figure}[htbp]
    \centering
    \includegraphics[width=.8 \columnwidth]{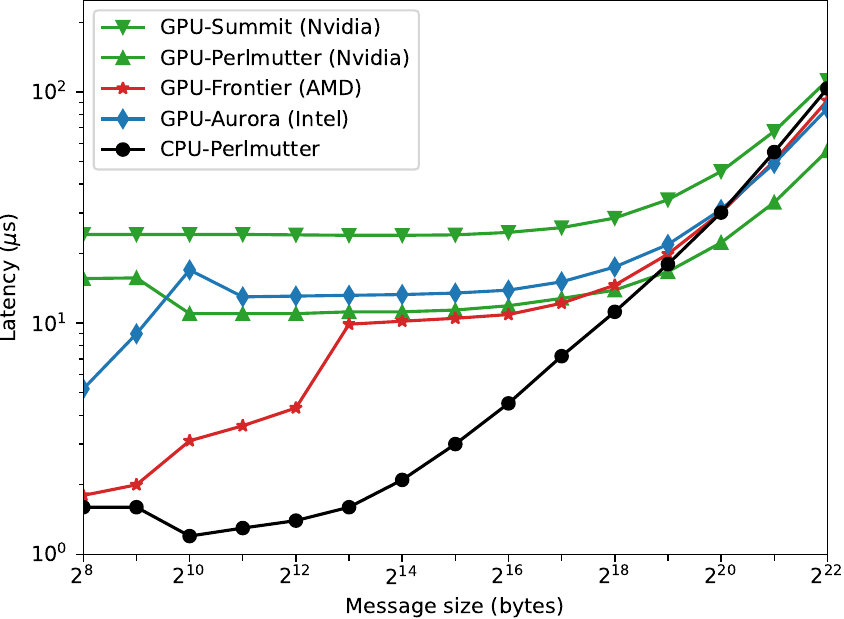}
    \caption{SF-pingpong test: MPI latency between two closest GPUs on the four (pre-)exascale machines, and between two CPU cores within a compute node on Perlmutter.
    Note the strikingly better performance on Frontier with small messages compared with other machines.}
    \label{fig:SF-Pingpong}
\end{figure}

\rev{
Although the experiments using NVSHMEM are promising, GPU-aware MPI is still PETSc's default communication mechanism on the GPU,
and we therefore designed the microbenchmark shown in Listing \ref{lst:sf-pingpong} to measure
its performance.} %
One of the star-forests used is illustrated in the right of Figure \ref{fig:SF},
where there are \texttt{n} consecutive roots on the first process and \texttt{n} consecutive leaves on the second.
The leaves are connected to the roots one-on-one in order.
The code works as if the two processes keep bouncing a message of \texttt{n*sizeof(double)} bytes to the other.
\begin{lstlisting}[caption={PetscSF code measuring message latency},label={lst:sf-pingpong},captionpos=b]
  MPI_Op op = MPI_REPLACE; // or MPI_SUM
  for (i=0; i<niter; i++) {
    PetscSFBcastBegin(sf,MPI_DOUBLE,rdata,ldata,op);
    PetscSFBcastEnd(sf,MPI_DOUBLE,rdata,ldata,op);
    PetscSFReduceBegin(sf,MPI_DOUBLE,ldata,rdata,op);
    PetscSFReduceEnd(sf,MPI_DOUBLE,ldata,rdata,op);
  }
\end{lstlisting}
When \texttt{op} is \texttt{MPI\_REPLACE}, the two buffers, \texttt{rdata} for root data and
\texttt{ldata} for leaf data, are directly
used as MPI send/receive buffers. The communication in this case is very similar to the
\textit{osu\_latency} test from the OSU Micro-Benchmarks \citep{OSUMicro}, with the exception
that we synchronize the device before sending messages for reasons mentioned in Section \ref{subsec:mpi-gpu-mismatch},
while \textit{osu\_latency} does not.
We named the test SF-pingpong and measured the one-way latency of a message, with
results on the four (pre-)exascale machines shown in Figure \ref{fig:SF-Pingpong}.
Note that on Frontier and Aurora, we treated the two GCDs (or tiles) within a GPU package as two separate GPUs.
In all tests, we used the two closest GPUs within a compute node, in other words, the two are either in the same package or connected
to the same CPU socket (as on Summit or Perlmutter).
For comparison, we also give the intra-node CPU-to-CPU message latency on Perlmutter measured with
the same code with host data.
From Figure \ref{fig:SF-Pingpong}, we can see that MPI latency with device data is generally much higher than that with host data
for small and medium messages, which implies GPU operations have a high start-up cost.
We can see that Perlmutter has much lower latency than Summit,
\rev{
presumably due to its newer hardware and software (see Table \ref{tab:machines}),
though we do not know enough about the workings of IBM Spectrum MPI to say exactly why.
}

\begin{figure}[htbp]
    \centering
    \includegraphics[width=.8 \columnwidth]{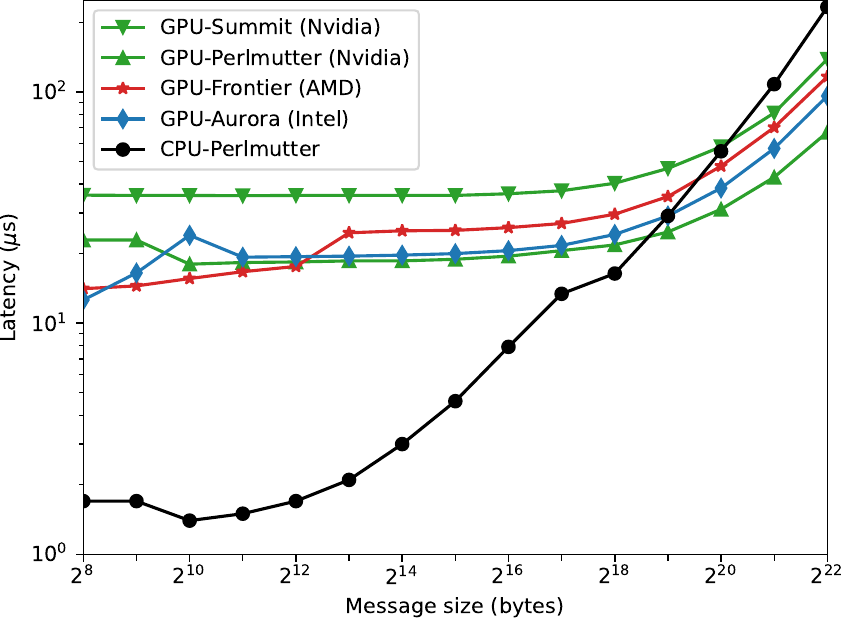}
    \caption{SF-unpack test: the test is similar to Figure \ref{fig:SF-Pingpong},
    except the latency contains
    the execution time of the unpack kernel after receiving data.
    Note the performance on Frontier will small messages did not stand out anymore as in Figure \ref{fig:SF-Pingpong}.
    Also note GPUs have much better performance than CPUs in the unpack kernel with big messages thanks to their higher memory bandwidth.
    }
    \label{fig:SF-Unpack}
\end{figure}

The GPU-Frontier performance in Figure \ref{fig:SF-Pingpong} stands out as its latency looks quite good for small messages, e.g., even as good as Perlmutter's CPU MPI latency for 256-byte messages.
Further investigation revealed that was misleading.
We changed the \texttt{op} in Listing \ref{lst:sf-pingpong} to \texttt{MPI\_SUM}, which
let the code add roots to leaves on broadcast and vice versa on reduce.
PETSc would allocate a buffer alongside the roots (leaves) and
call an unpack kernel to add values in the buffer to the roots (leaves).
We named the test SF-unpack and measured again the one way latency of a message, with results shown in Figure \ref{fig:SF-Unpack}.
We can see for small messages (therefore light unpack kernels) on Frontier, the latency increased dramatically from Figure \ref{fig:SF-Pingpong}.
Further analysis revealed that Cray-MPICH on Frontier would stage GPU messages on host for messages
smaller than a threshold and do host-to-host message passing instead,
while maintaining cache coherency between the CPU and GPU.
In the SF-pingpong test, we did not change the root (leaf) data, therefore the cached data on host was used in
subsequent iterations of message passing, giving an ultra-low average latency.
Meanwhile, in the SF-unpack test, the root (leaf) data was updated in each iteration, making host-caching useless.
In real applications, the data sent usually changes between iterations.
Therefore we deem the latency in Figure~\ref{fig:SF-Unpack} closer to the real
MPI latency for small messages on Frontier.
From the figure, we can also observe that GPUs have much better performance than
CPUs in the unpack kernel with big messages thanks to their higher memory bandwidth.

\section{Portable matrix assembly on GPUs} \label{sec:matassembly} %
In Section \ref{sec:petsc_port}, we previewed some new PETSc matrix assembly APIs for GPUs in the Jacobian computation.
In this section we discuss them in detail.
PETSc has a rich set of APIs for CPU (host) matrix assembly.
In the past, to achieve best performance users needed to call \texttt{MatXAIJSetPreallocation()} in advance to preallocate memory for the matrix, but the introduction of a hash table-based matrix assembly approach has rendered explicit preallocation no longer necessary in most cases.
Next,  users call \texttt{MatSetValues()} to insert blocks of values, e.g. element matrices from finite elements,
into the matrix using global indices. These values could be local (i.e., owned by the calling MPI process) or remote (i.e., owned by other processes), determined by the layout of the matrix.
For local values, PETSc directly inserts them, while for remote ones, PETSc might stash them until \texttt{MatAssemblyBegin()/End()} are called,
when MPI communication is used to distribute stashed values and the matrix is finally assembled.
While these APIs are convenient, they are all host APIs working with host memory, and can be called only by a single CPU thread.
Directly porting \texttt{MatSetValues()} and related functions to a GPU is not feasible, since if multiple GPU threads call  \texttt{MatSetValues()},
PETSc would need to stash remote values (and potentially allocate new stash area on device),  do binary searches to find locations to insert local values,
and use atomics to prevent data races when multiple GPU threads insert to the same non-zero location.
Though these operations keep memory utilization low, they are too latency-sensitive for efficient computing on GPUs.
Various approaches, like coloring to avoid atomics or using lookup tables to avoid binary search, were proposed \citep{cecka2011assembly, trotter2023targeting}.
But these were for a single process and, in effect, shifted the burden to application developers.
To support MPI parallelism and provide users convenience close to \texttt{MatSetValues()},
we designed a new set of coordinate-based (COO) matrix assembly APIs for matrices in the popular compressed sparse row (CSR) format.
In both PETSc's native and GPU formats, as well as Hypre's ParCSR,
MPI parallel matrices are distributed row-wise across MPI processes
with diagonal (intra-process coupling) and off-diagonal (inter-process coupling) blocks stored separately
as two sequential CSR matrices.
The classic COO format consists of three arrays, \texttt{i[]}, \texttt{j[]}, \texttt{v[]} of equal length,
in which the assembled matrix $A$ is defined as the sum of each contribution \texttt{v[k]} to entry $a_{\texttt{i[k]},\texttt{j[k]}}$,
so that each index $k$ represents a distinct nonzero.
In nonlinear and transient solves, one needs to repeatedly assemble matrices with the same nonzero pattern but different numeric values.

PETSc splits COO assembly into a symbolic preallocation stage with \texttt{MatSetPreallocationCOO(A,n,i,j)}, and
one or more numeric stages with \texttt{MatSetValuesCOO (A,v,mode)}.
During preallocation, it analyzes indices in \texttt{i/j[]} of length \texttt{n} on the host, exchanges information about remote entries,
finalizes the sparsity pattern of the diagonal and off-diagonal blocks,
preallocates device memory, and builds MPI communication plans.
The arrays \texttt{i/j[]} can be freed after this stage. We allow negative indices in \texttt{i/j[]}, meaning
the corresponding entries will be ignored, a convenient way to handle boundary conditions.
\texttt{MatSetValuesCOO()} sets elements of the matrix, where \texttt{v[]} is an array on the device that has the same length and follows
the same order as \texttt{i/j[]}.
Each entry (with non-negative indices) is destined for the owned diagonal, owned off-diagonal block, or a send buffer.
The implementation first calls a kernel to fill the send buffer on the device and initiates the GPU-aware MPI communication,
then calls two asynchronous kernels filling nonzeros in the diagonal and off-diagonal blocks, in which each thread accumulates into a single nonzero entry.
After completing communication, the implementation calls two similar kernels unpacking entries from the receive buffer.
With knowledge of the sparsity pattern and insertion order in advance, our implementation completely avoids data races and atomics with some helper data structures.

With PETSc's COO matrix assembly, users need to provide \texttt{i[]} and \texttt{j[]} at once.
In finite element assembly, they can loop over elements and precompute \texttt{i/j[]} on the host, assuming entries in the same element matrix will be
stored contiguously in the arrays.
Meanwhile, to prepare for the concurrent computing of element matrices on the device, one also needs to know the offset of each element in the array \texttt{v[]}.
With homogeneous elements so that each element matrix is the same size,
the offsets can be computed analytically. Otherwise, one can create an auxiliary array, say
\texttt{offset[]}, to store the information and also copy it to the device.
Then in the Jacobian, each GPU thread in charge of one element
conveniently gets a pointer by \texttt{\&v[offset[tid]]}, to store element matrix values, where \texttt{tid} is the thread id.

The above COO matrix assembly APIs are portable across host and device, so that the value array \texttt{v[]} can be on device or on the host,
depending on the type of the matrix.  For example, if $A$'s type is \texttt{MATAIJ}, a PETSc host CSR matrix type, then \texttt{v[]} needs to be on the host.
If it is \texttt{MATAIJCUSPARSE}, a PETSc matrix type for CUDA devices, then \texttt{v[]} can be host or device memory.
If it is \texttt{MATAIJKOKKOS}, a PETSc matrix type provided
by the PETSc/Kokkos backend, then \texttt{v[]} needs to be in the Kokkos default memory space.
To be complete, we also provide {\tt VecSetPreallocationCOO(x,n,i)} and
{\tt VecSetValuesCOO(x,v,mode)} for COO vector assembly on device.

\section{Asynchronous linear solvers on GPUs} \label{sec:asynchronous} %

\subsection{Distributed asynchronous CG with PetscSF over NVSHMEM} \label{subsec:cg_nvshmem}
In Section \ref{petscsf-nvshmem} we introduced a synchronization-free stream-aware PetscSF implementation with NVSHMEM.
Taking advantage of that, we were able to adapt CG, the conjugate gradient Krylov solver in PETSc,
to a prototype asynchronous version CGAsync. The full design was presented in \cite{PetscSF_TPDS_2021}.
CGAsync runs with multiple MPI ranks and GPUs, does all its computation and communication on device,
and does not need any synchronization on host.
With a modular design, the PETSc CG implementation contains calls to PETSc vector and matrix operations and MPI collectives.
In adapting CG to CGAsync, we had to make some changes.
Key implementation differences between CG and CGAsync include:
(1) A handful of PETSc routines they call are different. There are two categories.
The first includes routines with scalar output parameters, for example, vector dot product.
 CG calls \texttt{VecDot(Vec x, Vec y, double *a)} with \texttt{a} being a host address,
while CGAsync calls \texttt{VecDotAsync(Vec x, Vec y, double *a)} with \texttt{a}
being a device address.
In {\tt VecDot}, each process calls cuBLAS routines
to compute a partial dot product and then copies it back to the host, where it calls
{\tt MPI\_Allreduce} to get the final dot product and stores it in the host buffer.
Thus {\tt VecDot} synchronizes the host and the device. While in {\tt VecDotAsync},
once the partial dot product from cuBLAS is computed, each process calls a NVSHMEM reduction operation on PETSc's
default stream to get the final result and stores it in the device buffer.
The second category of differences includes routines with scalar input parameters,
such as \texttt{VecAXPY(Vec y,double a, Vec x)} calculating {\tt y += a*x}.
CG calls \texttt{VecAXPY} while CGAsync calls \texttt{VecAXPYAsync(Vec y,double *a,Vec x)} with
\texttt{a}
being a device pointer, so that \texttt{VecAXPYAsync} can be queued to a stream and \texttt{a} is computed on the device.
(2) CG does scalar arithmetic (e.g., divide two scalars) on the CPU, while CGAsync does them with
tiny \textit{scalar kernels} on the device.
(3) CG checks convergence (by comparison) in every iteration on the host to determine whether it should stop while
CGAsync does not.
Users need  to specify the maximum number of iterations; nevertheless, this could be improved by checking for convergence every few (e.g., 20) iterations.

\begin{figure}[htbp]
\begin{center}
\subfloat{\includegraphics[width=1.0\linewidth]{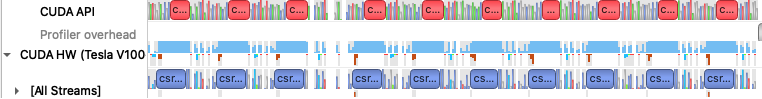}}\\
\vspace{10pt}
\subfloat{\includegraphics[width=1.0\linewidth]{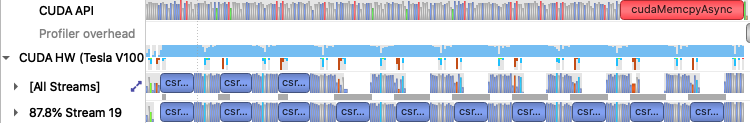}}
\caption{Timeline of CG (top) and CGAsync (bottom) on rank 2. Each ran ten iterations.
The blue \textit{csr...} bars are csrMV (i.e., SpMV) kernels in cuSPARSE, and the red \textit{c...} bars are
cudaMemcpyAsync() copying data from device to host.
}
\label{fig:CG}
\end{center}
\end{figure}

We tested CG and CGAsync without preconditioning on a single Summit compute node with two sparse matrices from
 the SuiteSparse Matrix Collection \citep{Florida}.
CG was run with PetscSF over the IBM Spectrum CUDA-aware MPI, and CGAsync was run with PetscSF over NVSHMEM.
The first matrix was Bump\_2911 with about 3M rows and 128M nonzero entries.
We ran both algorithms 10 iterations with 6 MPI ranks and one GPU per rank.
Fig. \ref{fig:CG} shows their timeline through the profiler NVIDIA NSight Systems.
The kernel launches (labeled \texttt{CUDA API}) in CG were spread over the
10 iterations. The reason was that in each iteration, there were multiple MPI calls
(mainly from distributed matrix-vector multiplication, vector dot and vector norm operations),
which constantly blocked the kernel launch
pipeline. In CGAsync, however, while the GPU was executing the 8th iteration (with profiling),
the CPU had launched \textit{all} kernels for the 10 iterations.
The long red bar {\tt cudaMemcpyAsync} indicates that after the kernel launches,  the CPU was idle, waiting for the final result from the GPU.

Test results show that the time per iteration for CG and CGAsync was about 690 $\mu s$ and 676 $\mu s$,
respectively. CGAsync gave merely a 2\% improvement.
This small improvement is because the matrix in question was huge, and computation took the vast majority of the time.
From profiling, we knew matrix-vector multiplication alone (excluding communication)
took 420 $\mu s$. If one removes the computational time, the improvement in communication time is substantial.
Unfortunately, because of bugs in the NVSHMEM library with multiple nodes,
we could not scale to more compute nodes.
Instead, we used a smaller matrix, Kuu from SuiteSparse, of about 7K rows and 340K nonzero entries to see how
CGAsync would perform in a strong-scaling sense. In the new tests,
time per iteration for CG and CGAsync was about 300 $\mu s$ and 250 $\mu s$.
CGAsync exhibited an improvement of 16.7\%.
Note that this improvement was achieved even though PetscSF/NVSHMEM had much higher message latency than
PetscSF/MPI (detailed in \cite{PetscSF_TPDS_2021} but not shown here).
Thus one can reasonably predict asynchronous solvers enabled by stream-aware communication have good potential.

\subsection{Improve solver asynchronicity with PetscDeviceContext and managed memory} \label{sec:PetscDeviceContext}

In addition to the ability to hide kernel launch latency, streams are useful for
achieving \emph{kernel concurrency}. Multiple kernels including memory copy operations
can be assigned to different streams, and these streams can run at the same time if there are sufficient resources.
Although using multiple streams in a confined scope might be manageable,
doing so at library level is hard, as one has to pass the stream information
through call chains, and maintain data dependence across streams.
In Section \ref{sec:challenges} we discussed the stream usage in PETSc.
To exploit kernel concurrency and manage asynchronicity in math library solvers,
we proposed a preliminary infrastructure for safe, seamless, and scalable integration of
asynchronous GPU streams in PETSc, detailed in \cite{faibussowitsch2023}.
The infrastructure consists of two new PETSc types, a C {\tt PetscDeviceContext} and a C++ {\tt Petsc::ManagedMemory} (\textit{device context} and  \textit{managed memory} thereafter).

Conceptually, a PETSc device context is an abstraction for vendor GPU streams \rev{ (not to be confused with 
the {\tt CUcontext} type in CUDA driver, which is an encapsulation for \textit{all} resources
and actions performed within the CUDA driver API).
It encodes the device ID over which it presides and the stream it manages, but also
bears other responsibilities.}
We provide APIs for one to create, synchronize, or fork and join device contexts.
With vendor programming models, the inter-stream data dependence is usually done via \textit{event} objects (e.g., {\tt cudaEvent\_t})
where a data producer records an event on a stream and a data consumer waits for the event on another stream.
But using event objects requires users to store or pass the objects around, \rev{rendering the library call chain error-prone.}
We now hide the complexity within PETSc device contexts and expose higher level context control APIs.
In addition, we provide memory registration APIs to assign memory regions a PETSc object ID. With that,
we can connect PETSc GPU {\tt Vec/Mat} objects and memory regions with device contexts.
It is done via a pair of Begin/End marking APIs that take arguments of  a device context, a PETSc object ID, and an access mode as below:

\begin{lstlisting}[label={lst:PetscDeviceContextMark},]
PetscDeviceContextMarkIntentFromIDBegin/End(PetscDeviceContext dctx, PetscObjectId obj, PetscMemoryAccessMode mode, const char descr[]);
\end{lstlisting}
They notify the runtime the PETSc object (or memory region) in the enclosed scope is accessed read-only, write-only or read-write.
Once an object has been marked, subsequent operations on it
are ordered \rev{to respect data dependency, while independent operations might be executed in parallel with multiple contexts. }
To enforce data dependency,
PETSc will internally create event objects for all input and output objects of an operation.
The latest write event or all the read events since
the last write are recorded after the operation has been launched; they are used as semaphores
for subsequent launches. If an object has previously recorded events, and the
new operation conflicts with a previous one, then PETSc will ensure that
the current operation does not begin until those events have completed. It is important to note that waiting for the
previous operation is done asynchronously on the device whenever possible.

\rev{
In PETSc, we often need scalars to 
mutate or scale larger objects or, in linear solvers, to
determine algorithm convergence. These scalars might be output by a kernel,
trivially modified or inspected, and then passed directly on to another kernel.
In the subsection above, we used raw device pointers to scalar variables,
resulting in non-portable code.
To overcome this, we designed {\tt Petsc::ManagedMemory},
which uses a \textit{std::future} for an opaque, host/device mirrored dual array of values.}
For example, its subclass {\tt ManagedReal} is used to manipulate real-valued scalar variables on device or host.
 {\tt ManagedReal} has the ability to symbolically
represent expressions, convert them into the corresponding CPU or GPU
kernels by using C++ expression templates, and then execute them on host or device, depending on runtime options. 
An example asynchronously normalizing a vector is shown in Listing \ref{lst:async_petsc}.

\begin{lstlisting}[caption={Asynchronous normalization of a PETSc vector {\tt v}},label={lst:async_petsc},captionpos=b]
PetscDeviceContext dctx;
Petsc::ManagedReal alpha; // subclass of ManagedMemory

// Retrieve the current device context,
PetscDeviceContextGetCurrentContext(&dctx);
// Asynchronously compute the norm on the GPU,
// storing its future value in alpha
VecNormAsync(v, NORM_2, &alpha, dctx);
// Evaluate the reciprocal expression asynchronously
alpha = Petsc::Eval(1.0 / alpha, dctx);
// Complete the normalization, asynchronously
VecScaleAsync(v, alpha, dctx);

// Materialize the value of alpha, copying device-to-host
// and synchronizing the stream transparently
PetscReal alpha_host = alpha;
printf("Value of reciprocal norm was %
\end{lstlisting}

\begin{figure}[htbp]
    \centering
    \includegraphics[width=.8 \columnwidth]{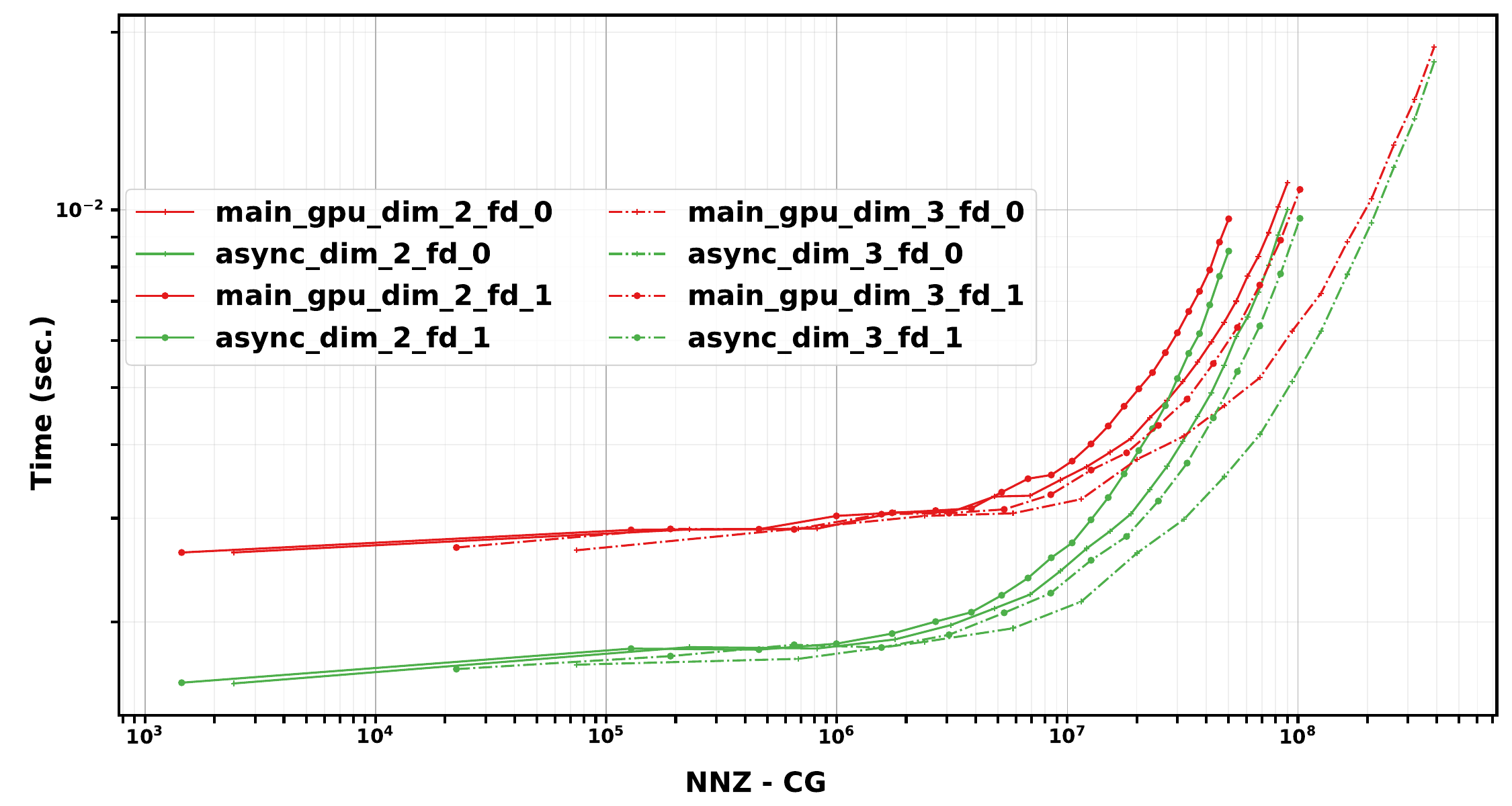}
    \includegraphics[width=.8 \columnwidth]{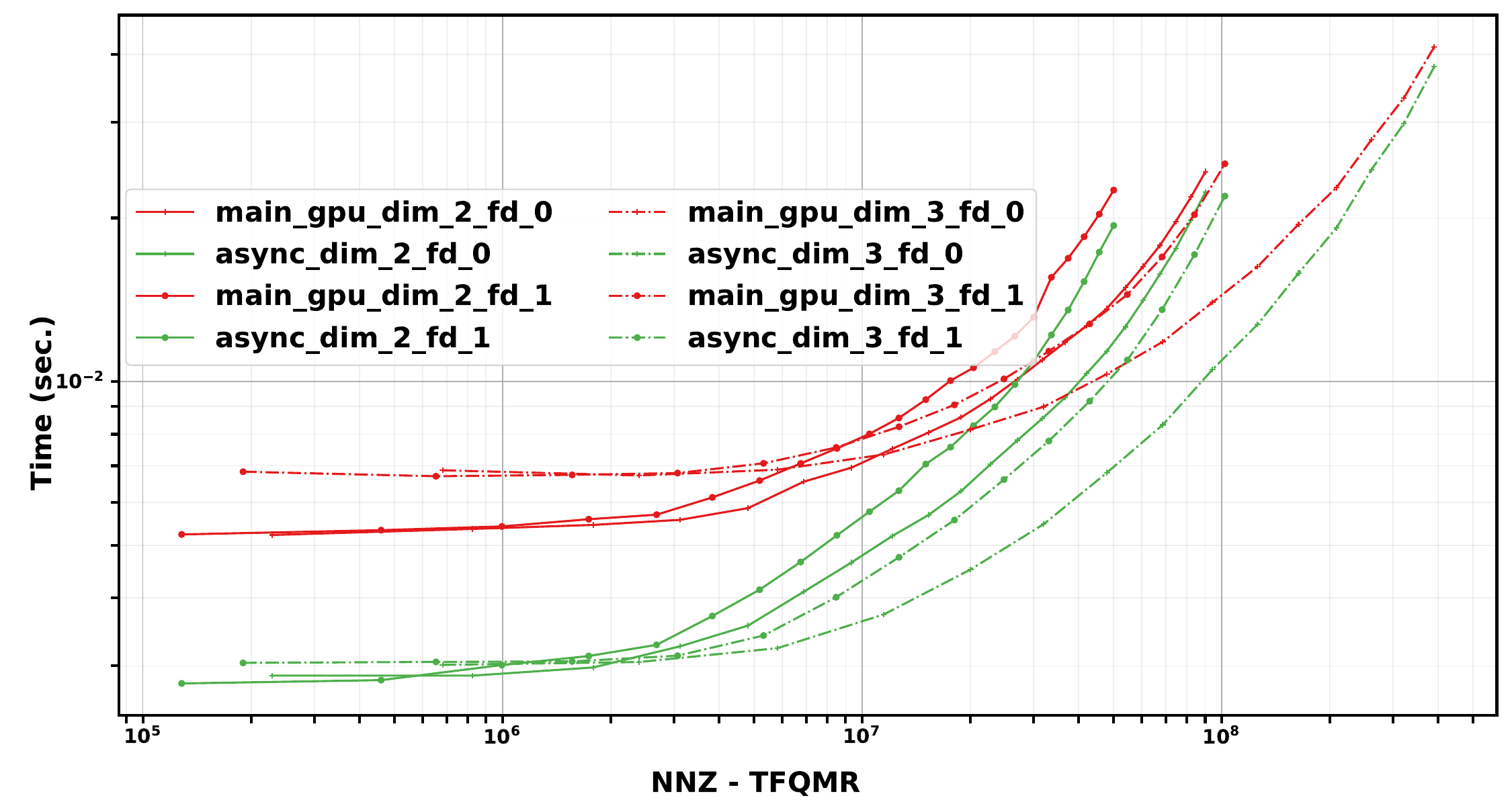}
    \caption{Synchronous vs asynchronous execution time of a solve versus the number of nonzeros of the system: CG (top) and TFQMR (bottom) }
    \label{fig:async_solve}
\end{figure}

With help of the PETSc device context and managed memory,
we implemented the CG solver and the Transpose-Free Quasi-Minimal Residual (TFQMR) solver \citep{freund1993tfqmr}, and tested on
one NVIDIA A100 GPU on the \textit{Polaris} computer at ALCF, a machine very similar to Perlmutter and employing identical GPUs.
We used a finite difference PDE (fd) problem to compare two configurations:
\begin{itemize}
    \item {\tt main\_gpu\_dim\_{\bf N}\_fd\_{\bf M}}: Synchronous solvers as the baseline. This is the default option in PETSc.
    \item {\tt async\_dim\_{\bf N}\_fd\_{\bf M}}: Asynchronous solvers implemented with PETSc device context and managed memory.
\end{itemize}
We used a 5/9-points (${\bf M}=0/1$) stencil for 2-D domains (${\bf N}=2$), and 9/27-points (${\bf M}=0/1$) stencil for 3-D domains (${\bf N}=3$).
The average solve time is shown in Figure \ref{fig:async_solve}, with respect to
the problem size represented by the number of nonzeros (NNZ) of the system.
The async implementation is faster than the baseline at every measured point,
starting from speed-ups in the order of $2$X for relatively small
problem sizes and with diminishing returns as the problem size increases,
when matrix-vector multiplication tends to dominate the overall runtime.
TFQMR shows a greater improvement than CG; for smaller problems (NNZ $\lesssim 10^6$)
it is nearly twice as fast versus only an 80\% improvement in CG.
In future work, we will integrate these newly implemented solvers with stream-aware PetscSF to exploit
asynchronous solvers in multi-node multi-GPU environments.

\section{Batched linear solvers on GPUs} \label{sec:batched_solvers} %

\begin{figure}[h!]
\begin{center}
\includegraphics[width=.8\columnwidth]{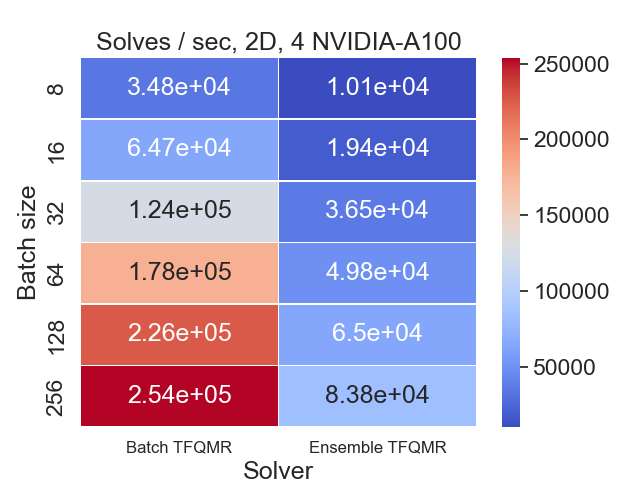}
\caption{Solver throughput vs problems size (batch size) for batched and ensemble solvers}
\label{fig:solver_heat_perlmutter2}
\end{center}
\end{figure}

\rev{Conceptually related (and complementary to) the use of asychronicity in linear solvers is the use of batching.}
{\it Batching}  is a technique for exposing PE-level parallelism in algorithms that previously ran on entire processes or multiple threads within a single MPI process.
Kinetic discretizations of magnetized plasmas, for example, advance the Vlasov-Maxwell system, which is then followed by a fully implicit time advance of a collision operator.
These collision advances are independent at each spatial point and are well suited to batch processing.
The full implicit time integrator in our Landau operator (Section \ref{sec:landau}) requires linear solves that can effectively run these many small systems on GPUs.
PETSc has developed batched version of two Krylov methods, TFQMR and BiCG, with diagonal preconditioning.
Batched iterative solvers of this form have also been deployed in the Kokkos-Kernels and Ginkgo libraries \citep{Liegeois2023,ginkgo-toms-2022}.
Figure \ref{fig:solver_heat_perlmutter2} shows the solves per second, as a function of batch size, of the batched and ``ensemble" solvers.
Ensemble solvers simply stack the individual systems in one large system and use standard PETSc solvers.
A 3X performance increase is observed with batching.
For details see \cite{adams2024performance}.

\section{Dense reformulations of algorithms} \label{sec:dense-formulate}

Reformulating algorithms to replace level-1 BLAS operations with level-2 or
level-3 operations generally improves performance,
but dense reformulations are more challenging to formulate for
flexible algorithms that use callbacks that cannot be vectorized.
In PETSc we have developed a dense reformulation of
the limited-memory BFGS method (L-BFGS, \cite{liu1989limited}), one of the most
popular optimization algorithms, that improves performance while
maintaining the flexibility of the original implementation.

At iteration $k$ of the L-BFGS algorithm to minimize $f(x)$, the two main steps are
(1) updating the approximation from $H_{k-1}$ to $H_k$, where $H_k$ is an approximation of the inverse Hessian $(\nabla f(x_k))^{-1}$, and (2) computing
the matrix-vector product $p = H_k g$.  The reference recursive algorithm for step~(2) consists of
$m$ rank-1 updates, followed by the application of a base inverse Hessian $H_0^{(k)}$,
followed by $m$ more rank-1 updates:

\vspace{\baselineskip}

\hrule
\removelatexerror
\begin{algorithm2e}[H]
\For{$j \in [k-1, \dots, k-m]$}{\label{line:rbfgsloop1start}
    $\alpha_j \gets (s_j^T g) / d_j$ \tcp*{dot($n$)}

    $g \gets g - \alpha_j y_j$ \tcp*{axpy($n$)}
}

$p \gets H_0^{(k)} g$ \tcp*{base operator}

\For{$j \in [k-m, \dots, k-1]$}{\label{line:rbfgsloop2start}
    $\beta_j \gets (y_j^T p) / d_j$ \tcp*{dot($n$)}

    $p \gets p + (\alpha_j - \beta_j) y_j$ \tcp*{axpy($n$)}
}
\Return{$p$}
\end{algorithm2e}
\hrule

\vspace{\baselineskip}
\noindent
The history size $m$ is fixed, and the
method requires the storage of history vectors $S_k = [ s_{k-m} | \cdots | s_{k-1}]$ and
$Y_k = [ y_{k-m} | \cdots | y_{k-1} ]$ and the precomputation of the dot products $d_j :=
s_j^T y_j$.
In the implementation of the recursive algorithm in PETSc, $H_0^{(k)}g$ is computed as a callback to an arbitrary linear operator.  This operator is not only arbitrary but may change from iteration to iteration, which allows this implementation of L-BFGS to support variable-metric methods \citep{davidon1991variable}. For many problems, variable-metric L-BFGS has been shown to converge faster than L-BFGS with a constant base operator $H_0^{(k)} \equiv H_0$ \citep{Dener2019}.
The recursive algorithm is work optimal, but it includes $2m$ synchronizations
(one for each dot product) and, as it is based on level-1 BLAS, has little temporal locality that takes advantage of the cache hierarchy.

A compact dense reformulation of L-BFGS was first studied by \cite{byrd1994representations}, which implements
$H_k g$ by adding to $H_0^{(k)}g$ the rank-$2m$ update
\[
\underbrace{\begin{bmatrix} -S_k R_k^{-T} & W_k  \end{bmatrix}}_{\Phi_k}
\begin{bmatrix}
    D_k + W_k^T Y_k & I \\
    I & 0
\end{bmatrix}
\underbrace{\begin{bmatrix}
    -R_k^{-1} S_k^T \\ W_k^T
\end{bmatrix}}_{\Phi_k^T} g,
\]
where $W_k = H_0^{(k)} Y_k$, $D_k = \mathrm{diag}(S_k^T Y_k)$, and $R_k = \mathrm{triu}(S_k^T Y_k)$.
This compact dense algorithm can be implemented with only one synchronization point (in computing $\Phi_k^T g$)
and all its flops are in dense matrix-vector products and triangular solves.  If $H_0$ is constant this compact dense algorithm requires only $O(m^2)$ more work than the
recursive algorithm, which is negligible when $n \gg m$.  When $H_0^{(k)}$ varies, however,
none of the vectors in $W_{k-1}$ can be reused as part of $W_k$, so $m-1$ additional matrix-vector products with $H_0^{(k)}$ are required to compute $W_k$ and $O(nm^2)$ additional work is required to compute $W_k^T Y_k$.

We have implemented an intermediate dense formulation of L-BFGS that computes $H_k g$ as
\begin{multline*}
H_k g =
\begin{bmatrix}
I & -S_k R_k^{-T}
\end{bmatrix} \\
\begin{bmatrix}
I & 0 \\
Y_k^T & I
\end{bmatrix}
\begin{bmatrix}
H_0^{(k)} & \\ & D_k
\end{bmatrix}
\begin{bmatrix}
I & Y_k \\ 0 & I
\end{bmatrix} \\
\begin{bmatrix}
I \\ -R_k^{-1} S_k^T
\end{bmatrix}
g.
\end{multline*}
Like the compact dense formulation, the product $H_k g$ is computed entirely with
dense matrix-vector products and triangular solves. This version has only two synchronization points (in the applications of $S_k^T$ and $Y_k^T$).  Unlike the compact dense formulation, there is no need for the vectors $W_k$ or the matrix $W_k^T Y_k$,
so $H_0^{(k)}$ can vary arbitrarily without increasing the work to update or apply $H_k$.

\begin{figure}[htbp]
    \centering
    \includegraphics[width=0.85 \columnwidth]{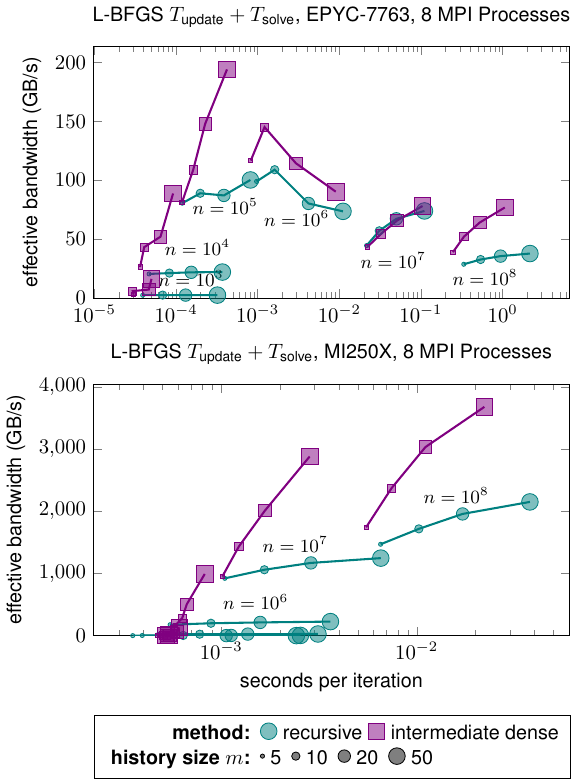}
    \caption{Comparison of L-BFGS implementations on one node of Frontier, using host CPUs (top) and GPUs (bottom). Marker size indicates the history size $m$ of the method, and measurements with the same problem size $n$ and different history sizes are connected.%
    }
    \label{fig:cdbfgs-frontier-8gpu}
\end{figure}
Figure~\ref{fig:cdbfgs-frontier-8gpu} presents a comparison of the recursive and
intermediate dense L-BFGS methods on one node of Frontier, using 8 MPI processes to drive either CPU-based computations (using 1 MPI rank per L3 cache) or GPU-based computations (using 1 MPI rank per GCD).  For different
problem sizes $n$ and different history sizes $m$ we plot
$T_{\text{update}} + T_{\text{solve}}$, the time it takes to update $H_{k-1}$ to $H_k$ and compute $p
= H_k g$.

Any algorithm implementing L-BFGS must, at minimum, read $S_k, Y_k, \in \mathbb{R}^{n
\times m}$, read $g \in \mathbb{R}^n$, and write $p \in \mathbb{R}^n$ at each
iteration, so we can define an {\it effective bandwidth} of each data point, $B_e := n(2m + 2)/(T_{\text{update}} + T_{\text{solve}})$, that we use as the y-axis in Figure~\ref{fig:cdbfgs-frontier-8gpu}.
This plot format is useful for comparing the tradeoffs between time-to-solution and efficiency for different algorithms and devices.

For small problem sizes ($n\leq 10^3$ on the CPUs, $n\leq 10^5$ on the GPUs),
when the latency of synchronization is the dominant cost, the intermediate dense
method exhibits runtimes that are almost independent of $m$, unlike the
recursive algorithm with its $2m$  synchronizations.
For the largest problems,
when a single vector does not fit into the last level of cache, a rank-1 update
cannot cache the vector that is being updated, so the $2m$ rank-1 updates of the
recursive method will require almost twice the memory traffic of two rank-$m$
updates of the intermediate dense method. We see this in the effective bandwidth
for the largest problem sizes on both CPUs and GPUs, and the same reasoning
explains why the intermediate dense method achieves almost twice the bandwidth
of the recursive dense method when all of the problem data fits into the last level of cache for CPU computations, such as
in Figure~\ref{fig:cdbfgs-frontier-8gpu} (top) when $n=10^5$ and $m=50$.
This data shows that the intermediate dense L-BFGS formulation in PETSc exhibits almost uniformly superior performance to the recursive formulation while retaining all of the flexibility of the recursive formulation in supporting variable-metric methods. \rev{We plan to study the benefits of these improved performances for training deep neural networks \citep{Zampini_PASC}.} Further dense reformulations of iterative methods that retain flexibility
are planned as future improvements to PETSc.

\section{\rev{Mini-application study 1: Solving 3D Laplacian with algebraic multigrid}} \label{sec:gamg}

In this section we demonstrate the performance portability of the 
PETSc/Kokkos backend over CUDA and HIP with a scaling study of the PETSc's built-in algebraic multigrid (AMG) solver {\tt PCGAMG}.
{\tt PCGAMG} uses the PETSc multigrid {\tt PCMG} framework and can thus
take advantage of optimized backend operations.
This ability to abstract the AMG algorithm with standard sparse linear algebra has facilitated its widespread use in the PETSc and wider computational science communities.
PETSc's built-in functionality for finite elements is used to discretize the Laplacian operator with second-order elements.
Each MPI process has a logical cube of hexahedral cells, with 64 such processes per node (e.g., 16 MPI tasks per A100).
Increasingly larger grids are generated by uniform refinements.

\begin{figure}[htbp]
\begin{center}
\includegraphics[width=.49\linewidth]{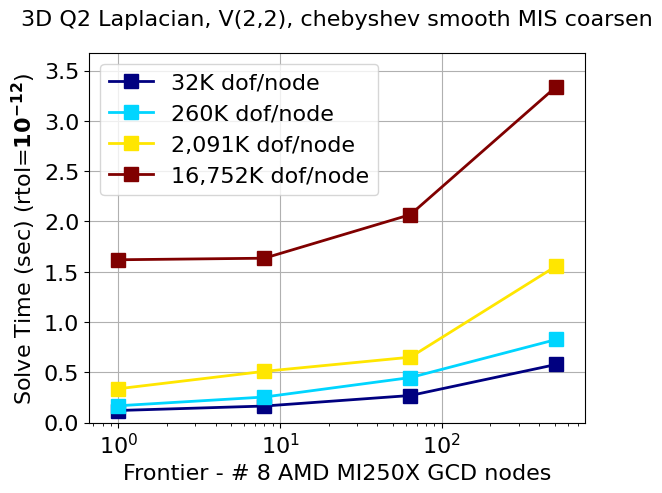}
\includegraphics[width=.49\linewidth]{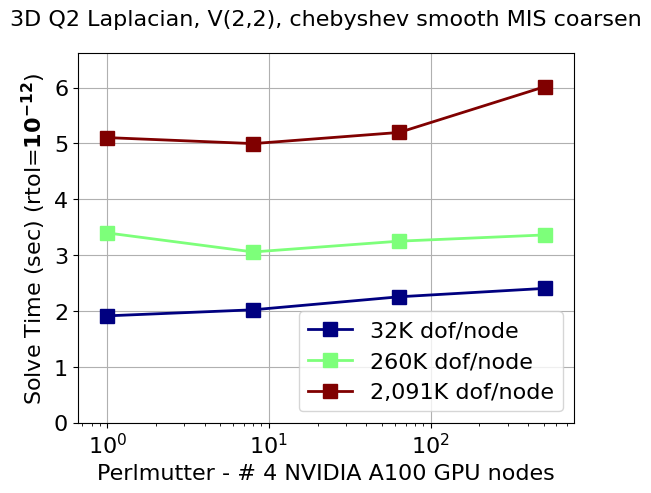}
\caption{Solve time (sec.) for 1 solve of a 3D Laplacian with Q2 elements, a relative residual tolerance of $10^{-12}$: Frontier (left) and Perlmutter (right).}
\label{fig:gamg_weak_scaling2}
\end{center}
\end{figure}

Figure \ref{fig:gamg_weak_scaling2} shows performance data for the solve phase with several subdomain sizes as a function of the number of nodes, keeping the same number of cells per MPI task, that is, weak scaling where horizontal lines are perfect, on Perlmutter and Frontier.
This shows that MPI parallel scaling is fairly good (there is a slight increase in iteration counts that is folded into the inefficiency) because the lines are almost flat, up to 512 nodes.

\begin{figure}[htbp]
\begin{center}
\includegraphics[width=.49\linewidth]{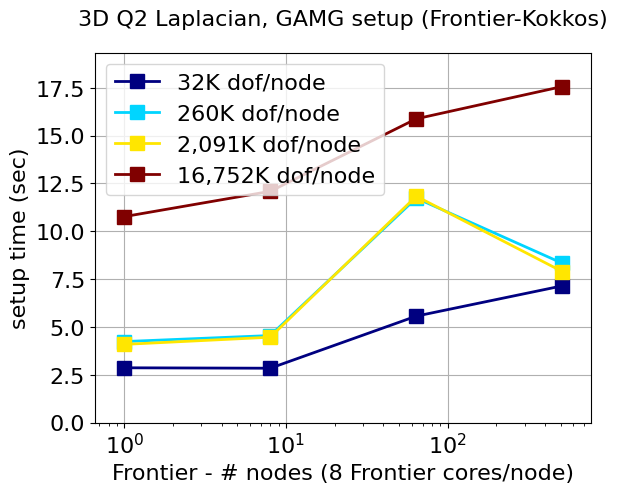}
\includegraphics[width=.49\linewidth]{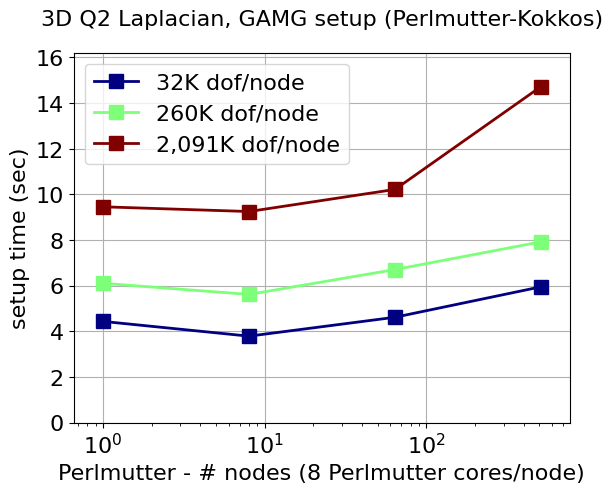}
\caption{Setup time (sec.) for 1 solve of a 3D Laplacian with Q2 elements: Frontier (left) and Perlmutter (right).}
\label{fig:gamg_weak_scaling_setup2}
\end{center}
\end{figure}

The ``setup" phase of AMG consists of two parts: the ``mesh" setup, which constructs the coarse grid spaces (mostly graph work), and the ``matrix" setup that constructs the coarse grid operators (a sparse matrix triple product).
The mesh setup phase is fully amortized for long simulations without adaptive mesh refinement; mesh setup is required for each new mesh.
The matrix setup is amortized for linear or secant Newton types of algorithms where the matrix does not change.
Matrix setup can sometimes be further amortized by lagging, where coarse grid operators are used from old fine grid operators.
Figure \ref{fig:gamg_weak_scaling_setup2} shows performance data for the setup phase with several subdomain sizes as a function of the number of nodes, keeping the same number of cells per MPI task. %
This data was generated with a test harness\footnote{\path{petsc/src/snes/tests/ex13.c}} that uses a two-level partitioning of Cartesian grids in {\tt DM}, first to nodes and then to MPI tasks on each node.
These experiments were run with 64 processes per node.

The matrix setup, mostly the sparse matrix triple product $P^TAP$ (or $RAP$) construction of the coarse grid operator ($A^{coarse}_{IJ} = P_{iI}A^{fine}_{ij}P_{jJ}$),
 has relatively high arithmetic intensity and a high degree of parallelism, but
is a challenge to optimize for GPUs.
The PETSc/Kokkos backend supports effective implementations of this operator.
We use the Kokkos-Kernels spgemm (sparse matrix-matrix multiplication) interface wrapping
around vendors' implementations (e.g., cuSparse, rocSparse)
as the building block within an MPI process, and use
PetscSF to carry out the complex communication between processes.

\section{\rev{Mini-application study 2: Time evolution of Landau collision integral}} \label{sec:landau} %

\begin{figure}[htbp]
    \centering
    \includegraphics[width=.94 \columnwidth]{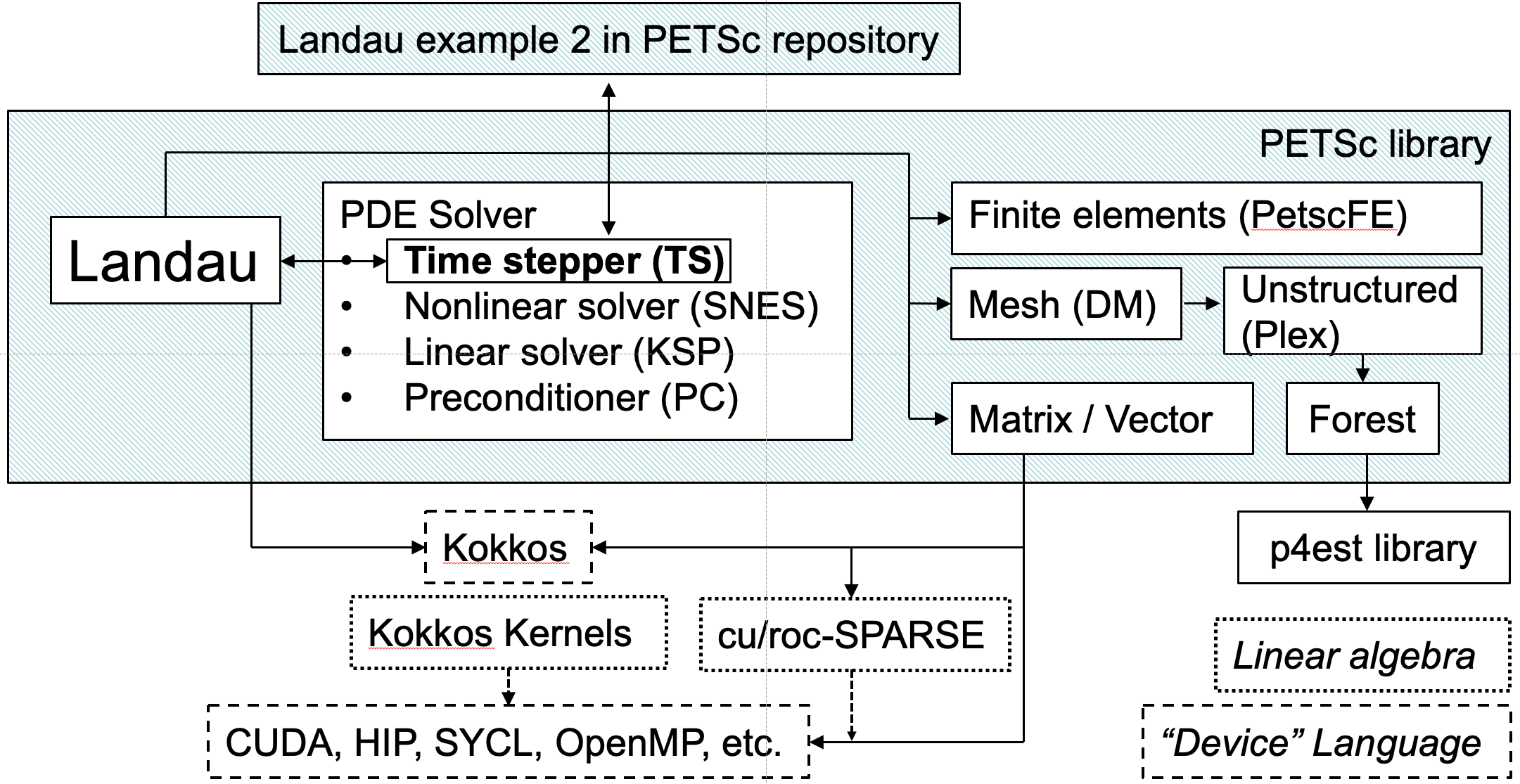}
    \caption{Code architecture of Landau collision application}
    \label{fig:lanau-petsc}
\end{figure}

Using PETSc's GPU support for the entire PDE solver stack, from time integrators to nonlinear solvers, batched linear solvers, and COO matrix assembly,
a fully GPU enabled Landau collision time advance is implemented with ``mini-app" drivers as a PETSc example \citep{Hirvijoki2017,AdamsHirvijokiKnepleyBrownIsaacMills2017,Adams2022a,adams2024performance}.
The structure of this code is shown in Figure \ref{fig:lanau-petsc}.
This method is an example of building an entire HPC PDE solver in PETSc, with a small driver code\footnote{  \path{petsc/src/ts/utils/dmplexlandau/tutorials/ex2.c}} mimicking an application, with verification tests, used for performance experiments, and a specific PDE operator (``Landau") that would, in general, be in user code.

\subsection{Landau collision integral}
\begin{figure}[h!]
    \centering
    \begin{subfigure}{.15\textwidth}
        \includegraphics[width=\textwidth]{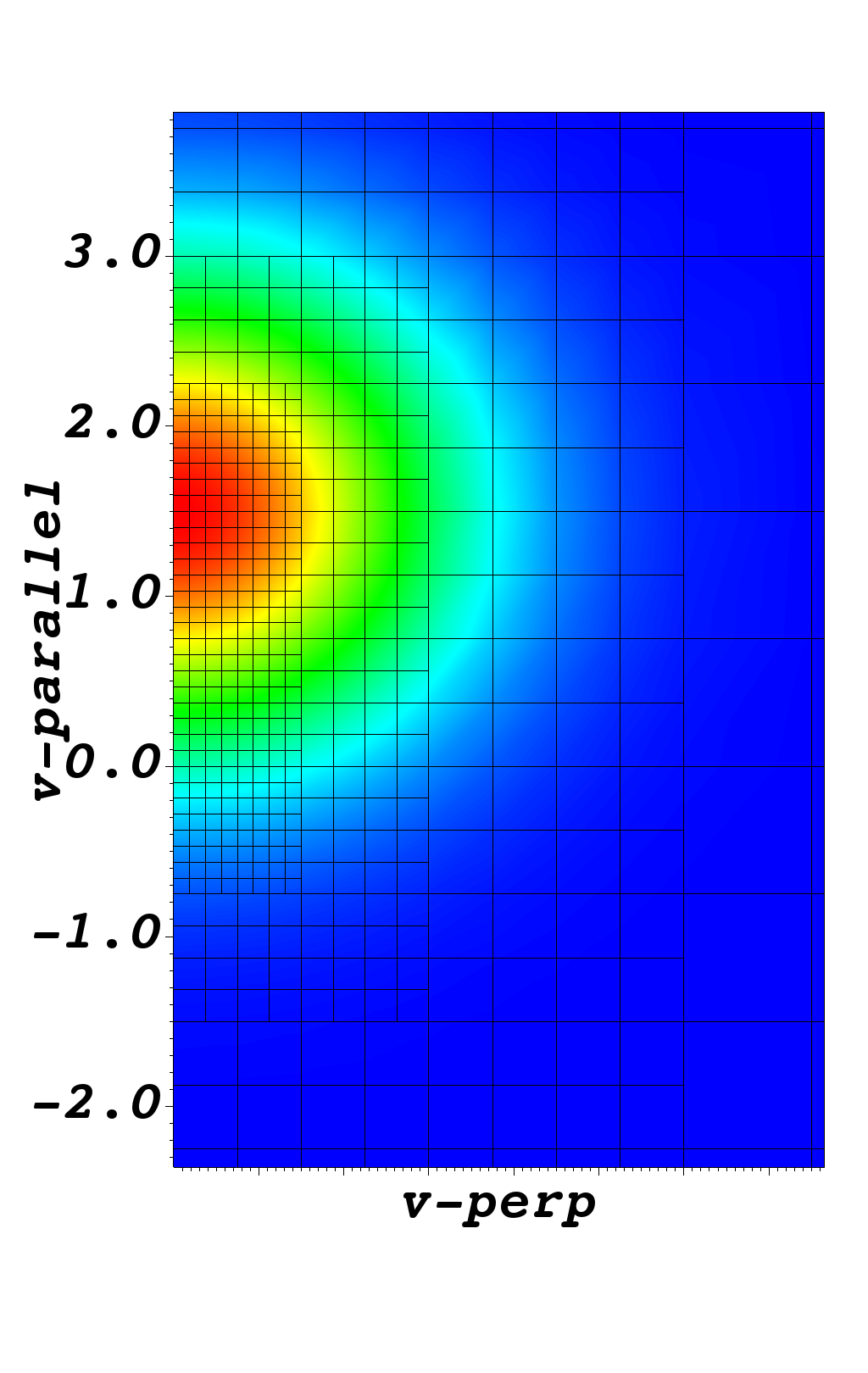}
        \caption{$t=0$} \label{fig:shifted_e0}
    \end{subfigure}
    \begin{subfigure}{.15\textwidth}
        \includegraphics[width=\textwidth]{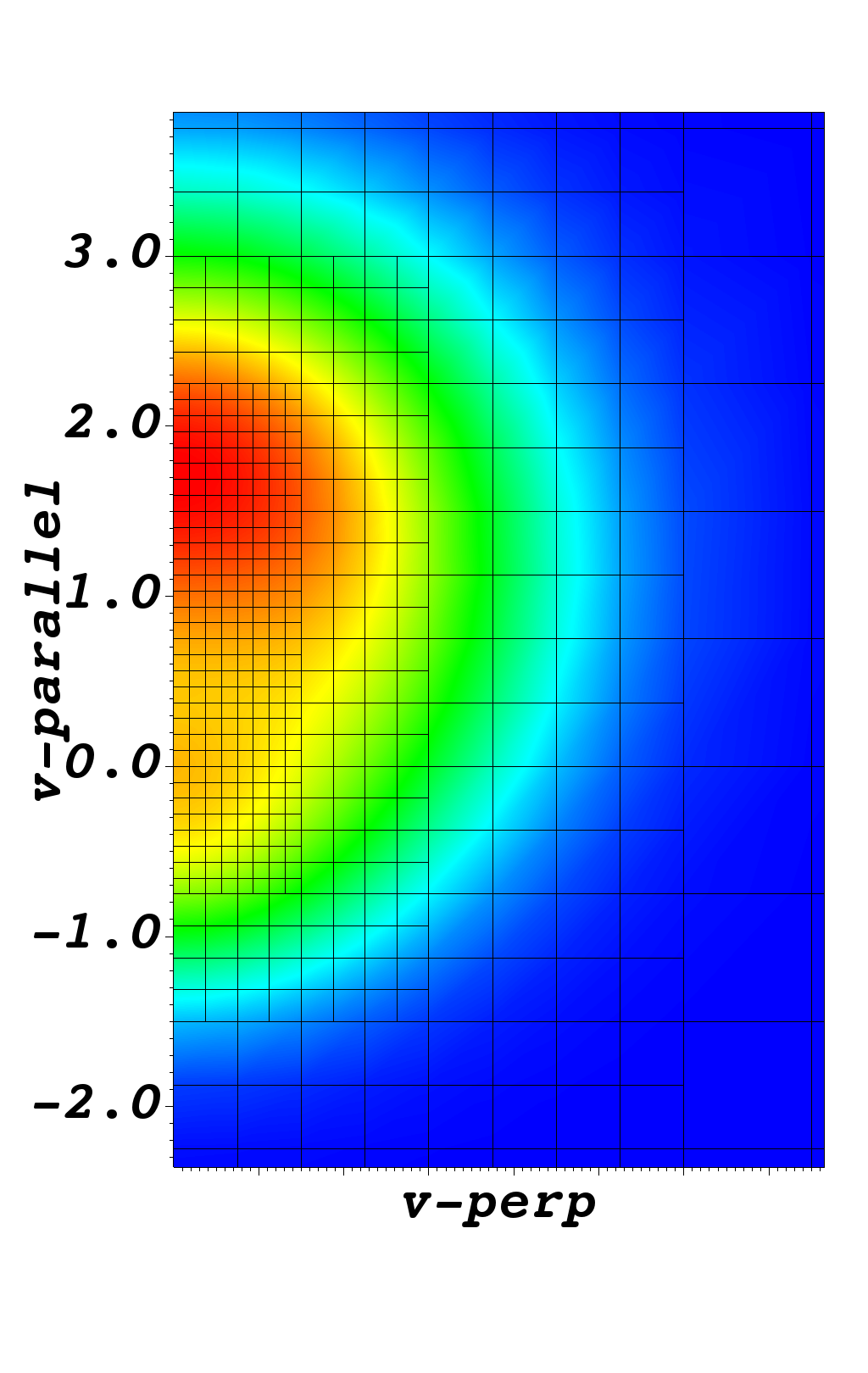}
        \caption{bi-modal} \label{fig:shifted_e2}
    \end{subfigure}
    \begin{subfigure}{.15\textwidth}
        \includegraphics[width=\textwidth]{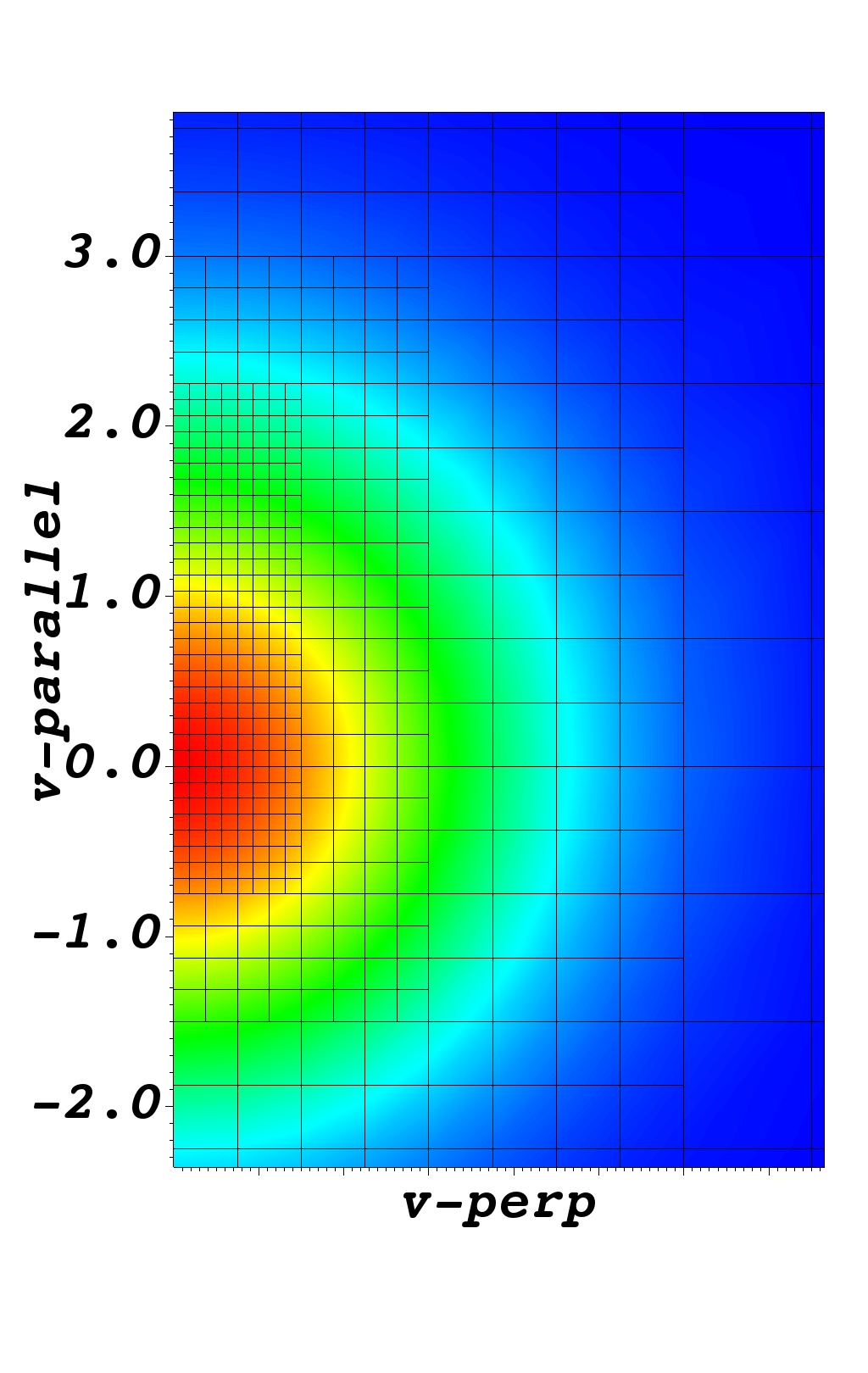}
        \caption{near equilibrium} \label{fig:shifted_e4}
    \end{subfigure}
\caption{Electron distributions of shifted Maxwellian deuterium plasmas: (a) initial condition, (b) penumbra in shift and early Maxwellian population, (c) near full thermalization with ions at origin (ions not shown)}
\label{fig:meshes}
\end{figure}

Many problems in physics are described with phase space models where density is a function of both space ($\mathbf{x}$) and velocity space ($\mathbf{v}$).
One such problem of interest to the DOE is that of magnetized plasmas for fusion energy science (FES), which are similar to several problems in astrophysics.
The governing equations for magnetized plasmas are the Vlasov-Maxwell-Landau (VML) system, where each species $\alpha$ (electrons and ions) are evolved according to
\begin{equation*}
\frac{df_\alpha}{dt} \equiv
\frac{\partial f_\alpha}{\partial t} + \frac{\partial  \vec{x}}{\partial t}
\cdot \nabla_x f_\alpha + \frac{\partial \vec{v}}{\partial t} \cdot \nabla_v f_\alpha = \sum_{\beta} C\left[f_\alpha,f_\beta\right]_{\alpha\beta} .
\end{equation*}
This equation is composed of the symplectic {\it Vlasov-Maxwell} term $\frac{df_\alpha}{dt}=0$ and a metric, or diffusive, collision operator $C$.
The Landau form of Fokker-Planck collisions is a velocity space operator and is the gold standard for fusion plasmas.
PETSc includes examples, with verification tests, that use our Landau collision operator to evolve $\frac{df_\alpha}{dt} = \sum_{\beta} C\left[f_\alpha,f_\beta\right]_{\alpha\beta}$, that runs an entire PDE simulation on the GPU with the PETSc GPU backends \citep{adams2024performance}.
Figure \ref{fig:meshes} shows the thermalization of a shifted Maxwellian distribution, with an initial drift velocity of $1.5$, with a stationary ion population (not shown). The large mass of ions results in small equilibrium velocity shown in Figure \ref{fig:shifted_e4}.

\subsection{Validation of Landau collision integral}

\begin{figure}[htbp]
\begin{center}
\includegraphics[scale=.45]{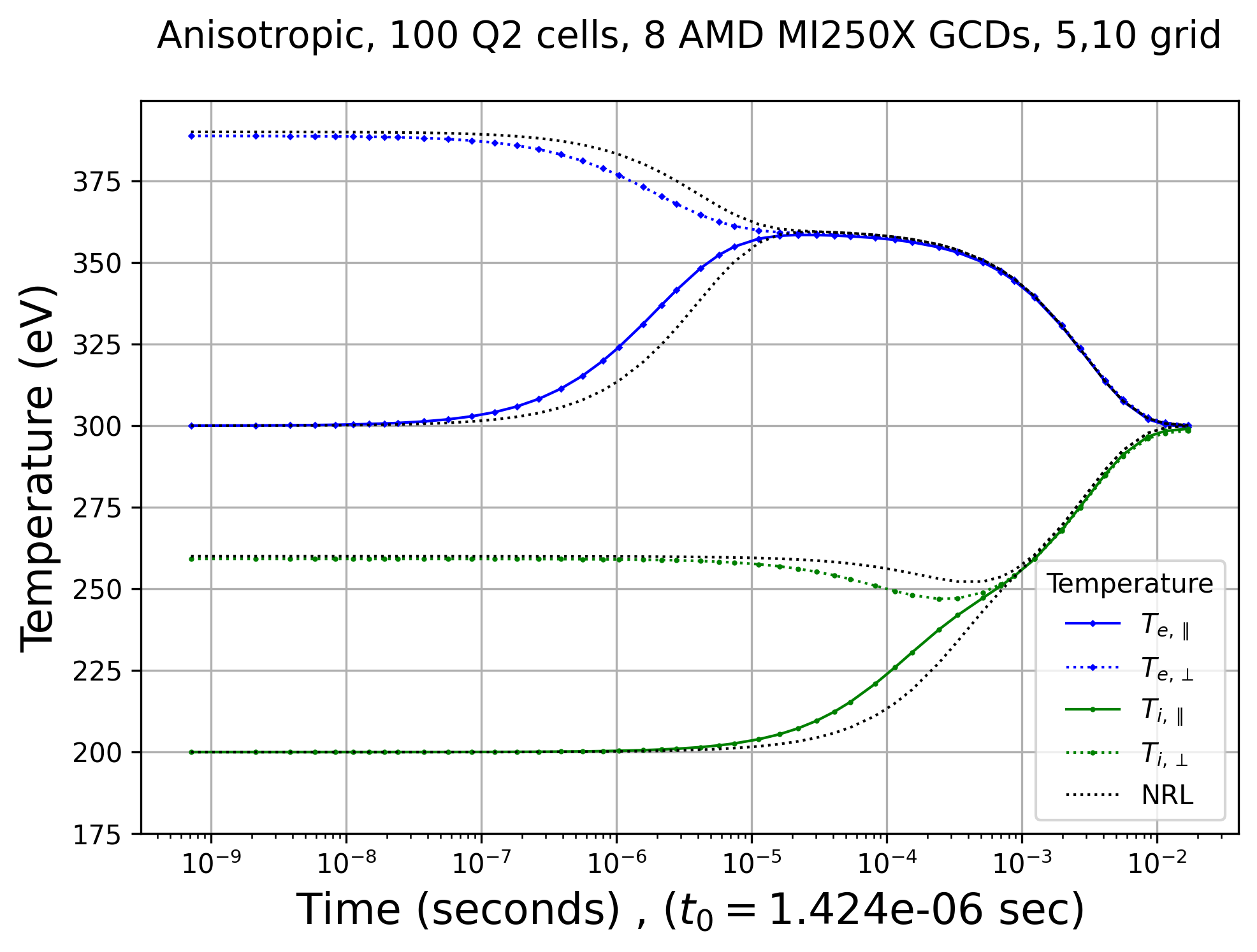}
\caption{Anisotropic relaxation test temperature vs. time, with the normalization time $t_0$, of the $P2$ element case, plotted with an analytical NRL results}
\label{fig:temperature-history}
\end{center}
\end{figure}

A common anisotropic plasma verification test has recently been published that initializes a two-species plasma (electrons and deuterium) with (4) different temperatures for the parallel and perpendicular temperature of each species that are allowed to evolve toward equilibrium.
A driver code runs many of these problems simultaneously to mimic its use in an application where 1,000s of spatial vertices would be processed simultaneously.
This problem uses PETc's adaptive time stepping with over 14,000 time steps to achieve near full thermalization, using new high-order simplex finite elements and verification with analytical results of thermalization rates \citep{adams2024performance}.
Figure \ref{fig:temperature-history} shows a temperature history for the $P2$ case.
Note, the difference with the NRL Plasma Formulary model are due to the NRL rates being derived with a simplifying assumption and others have observed similar differences \citep{Hager2016}.

\section{Conclusion and Vision for the Future} \label{sec:conclusion} %
We have summarized some of the challenges encountered and the advances made toward providing
performance-portable support for GPUs in PETSc over the course of ECP.
This effort has required development ranging across the different levels of the PETSc software stack.
As accelerated GPU computation continues to shape the HPC field,
we will continue to improve PETSc's capabilities for this, adding new features (e.g., support for the block CSR matrix format on GPUs), and offloading more computations.
We hope to consolidate PETSc GPU backends to simplify implementation and maintenance, and
we will continue collaborating with developers of PETSc external libraries to smooth the interface for GPU data passing between libraries.
An important direction that PETSc developers are currently pursuing is adding robust support for machine learning computations in PETSc
to support a broad range of applications, which demands high GPU performance.

Maintaining a stable interface is important for a numerical library like PETSc, but our experience has shown the tension between that goal and the goal of portable performance for pre- and early exascale machines.  In some cases, as in the dense reformulation of L-BFGS, additional parallelism can be found within existing interfaces that is portable across systems; in other cases, as in the introduction of the {\tt ManagedMemory} type, something that is technically a change to the library's interface
finds additional parallelism within existing programming patterns used by applications.
First in introducing {\tt PetscSF} and now in {\tt MatSetValuesCOO()}, PETSc has had success with new interfaces that (a) are declarative, (b) expose as much of the parallelism within the desired computation as possible, and (c) are independent of the computational resources used to execute the operation and the memory resources holding the arguments.  Once introduced, interfaces with these properties appear to become stable features of the library, and applications that adopt them should see benefits.
In the future, the continued evolution of PETSc along these lines will help applications
achieve portable performance for larger and larger percentages of their workflows, both on exascale systems and those that will follow.

\begin{acks}
The authors were supported by the U.S. Department of Energy, Office of Science, Advanced Scientific Computing Research under Contract DE-AC02-06CH11357. This research used resources of the Argonne Leadership Computing Facility, which is a DOE Office of Science User Facility supported under Contract DE-AC02-06CH11357,
resources of the National Energy Research Scientific Computing Center,
a Department of Energy Office of Science User Facility, %
and resources of the Oak Ridge Leadership Computing Facility at the Oak Ridge National Laboratory,
which is supported by the Office of Science of the U.S. Department of Energy under Contract No. DE-AC05-00OR22725.
The work was partially done on a pre-production supercomputer with early versions of the Aurora software development kit.
The authors were partially supported by the Exascale Computing Project (17-SC-20-SC), a collaborative effort of the U.S. Department of Energy Office of Science and the National Nuclear Security Administration. MGK was partially supported by NSF CSSI award 1931524. SZ acknowledges the support of the Extreme Computing Research Center, King Abdullah University of Science and Technology.

\end{acks}

\bibliographystyle{SageH}
\bibliography{mybibfile}

\begin{thebibliography}{43}
\providecommand{\natexlab}[1]{#1}
\providecommand{\url}[1]{\texttt{#1}}
\providecommand{\urlprefix}{URL }
\expandafter\ifx\csname urlstyle\endcsname\relax
  \providecommand{\doi}[1]{DOI:\discretionary{}{}{}#1}\else
  \providecommand{\doi}{DOI:\discretionary{}{}{}\begingroup \urlstyle{rm}\Url}\fi

\bibitem[{Adams et~al.(2022{\natexlab{a}})Adams, Balay, Marin, McInnes, Mills, Munson, Zhang, Zhang, Brown, Eijkhout, Faibussowitsch, Knepley, Kong, Kruger, Sanan, Smith and Zhang}]{petsc-community2022}
Adams M, Balay S, Marin O, McInnes LC, Mills RT, Munson T, Zhang H, Zhang J, Brown J, Eijkhout V, Faibussowitsch J, Knepley M, Kong F, Kruger S, Sanan P, Smith BF and Zhang H (2022{\natexlab{a}}) The {PETSc} community as infrastructure.
\newblock \emph{IEEE CiSE} 24(3): 6--15.
\newblock \doi{10.1109/MCSE.2022.3169974}.

\bibitem[{Adams et~al.(2022{\natexlab{b}})Adams, Brennan, Knepley and Wang}]{Adams2022a}
Adams MF, Brennan DP, Knepley MG and Wang P (2022{\natexlab{b}}) Landau collision operator in the {CUDA} programming model applied to thermal quench plasmas.
\newblock In: \emph{2022 IEEE International Parallel and Distributed Processing Symposium (IPDPS)}. pp. 115--123.
\newblock \doi{10.1109/IPDPS53621.2022.00020}.

\bibitem[{Adams et~al.(2017)Adams, Hirvijoki, Knepley, Brown, Isaac and Mills}]{AdamsHirvijokiKnepleyBrownIsaacMills2017}
Adams MF, Hirvijoki E, Knepley MG, Brown J, Isaac T and Mills R (2017) Landau collision integral solver with adaptive mesh refinement on emerging architectures.
\newblock \emph{SIAM Journal on Scientific Computing} 39(6): C452--C465.
\newblock \doi{10.1137/17M1118828}.

\bibitem[{Adams et~al.(2024)Adams, Wang, Merson, Huck and Knepley}]{adams2024performance}
Adams MF, Wang P, Merson J, Huck K and Knepley MG (2024) A performance portable, fully implicit landau collision operator with batched linear solvers.
\newblock Summited to SISC.

\bibitem[{{AMD}(2024)}]{HIP}
{AMD} (2024) {HIP} programming manual.
\newblock \urlprefix\url{https://rocm.docs.amd.com/projects/HIP/en/latest/user_guide/programming_manual.html}.

\bibitem[{Amestoy et~al.(2001)Amestoy, Duff, L'Excellent and Koster}]{mumps01}
Amestoy PR, Duff IS, L'Excellent JY and Koster J (2001) A fully asynchronous multifrontal solver using distributed dynamic scheduling.
\newblock \emph{SIAM Journal on Matrix Analysis and Applications} 23(1): 15--41.

\bibitem[{Anzt et~al.(2022)Anzt, Cojean, Flegar, Göbel, Grützmacher, Nayak, Ribizel, Tsai and Quintana-Ortí}]{ginkgo-toms-2022}
Anzt H, Cojean T, Flegar G, Göbel F, Grützmacher T, Nayak P, Ribizel T, Tsai YM and Quintana-Ortí ES (2022) {Ginkgo: A Modern Linear Operator Algebra Framework for High Performance Computing}.
\newblock \emph{ACM Transactions on Mathematical Software} 48(1): 2:1--2:33.
\newblock \doi{10.1145/3480935}.

\bibitem[{Balay et~al.(2024)Balay, Abhyankar, Adams, Brown, Brune, Buschelman, Dalcin, Dener, Eijkhout, Gropp, Karpeyev, Kaushik, Knepley, May, McInnes, Mills, Munson, Rupp, Sanan, Smith, Zampini, Zhang, Zhang and Zhang}]{petsc-3.21}
Balay S, Abhyankar S, Adams MF, Brown J, Brune P, Buschelman K, Dalcin L, Dener A, Eijkhout V, Gropp WD, Karpeyev D, Kaushik D, Knepley MG, May DA, McInnes LC, Mills RT, Munson T, Rupp K, Sanan P, Smith BF, Zampini S, Zhang H, Zhang H and Zhang J (2024) {PETS}c users manual.
\newblock Technical Report ANL-21/39 - Revision 3.21, Argonne National Laboratory.
\newblock \urlprefix\url{https://petsc.org}.

\bibitem[{Beckingsale et~al.(2019)Beckingsale, Burmark, Hornung, Jones, Killian, Kunen, Pearce, Robinson, Ryujin and Scogland}]{RAJA}
Beckingsale DA, Burmark J, Hornung R, Jones H, Killian W, Kunen AJ, Pearce O, Robinson P, Ryujin BS and Scogland TR (2019) {RAJA}: Portable performance for large-scale scientific applications.
\newblock In: \emph{2019 IEEE/ACM International Workshop on Performance, Portability and Productivity in HPC (P3HPC)}. IEEE, pp. 71--81.

\bibitem[{Bertoni et~al.(2020)Bertoni, Kwack, Applencourt, Ghadar, Homerding, Knight, Videau, Zheng, Morozov and Parker}]{colleen2020}
Bertoni C, Kwack J, Applencourt T, Ghadar Y, Homerding B, Knight C, Videau B, Zheng H, Morozov V and Parker S (2020) Performance portability evaluation of opencl benchmarks across intel and nvidia platforms.
\newblock In: \emph{2020 IEEE International Parallel and Distributed Processing Symposium Workshops (IPDPSW)}. pp. 330--339.
\newblock \doi{10.1109/IPDPSW50202.2020.00067}.

\bibitem[{Brown et~al.(2012)Brown, Knepley, May, McInnes and Smith}]{bkmms2012}
Brown J, Knepley MG, May DA, McInnes LC and Smith BF (2012) Composable linear solvers for multiphysics.
\newblock In: \emph{Proceeedings of the 11th {International Symposium on Parallel and Distributed Computing} ({ISPDC} 2012)}. IEEE Computer Society, pp. 55--62.

\bibitem[{Byrd et~al.(1994)Byrd, Nocedal and Schnabel}]{byrd1994representations}
Byrd RH, Nocedal J and Schnabel RB (1994) Representations of quasi-newton matrices and their use in limited memory methods.
\newblock \emph{Mathematical Programming} 63(1-3): 129--156.

\bibitem[{Cecka et~al.(2011)Cecka, Lew and Darve}]{cecka2011assembly}
Cecka C, Lew AJ and Darve E (2011) Assembly of finite element methods on graphics processors.
\newblock \emph{International Journal for Numerical Methods in Engineering} 85(5): 640--669.

\bibitem[{Davidon(1991)}]{davidon1991variable}
Davidon WC (1991) Variable metric method for minimization.
\newblock \emph{SIAM Journal on optimization} 1(1): 1--17.

\bibitem[{Davis and Hu(2011)}]{Florida}
Davis TA and Hu Y (2011) The university of florida sparse matrix collection.
\newblock \emph{ACM Transactions on Mathematical Software (TOMS)} 38(1): 1--25.

\bibitem[{Demmel et~al.(2024)Demmel, Gilbert and Li}]{superlu}
Demmel J, Gilbert J and Li X (2024) {SuperLU Github}.
\newblock \urlprefix\url{https://github.com/xiaoyeli/superlu}.

\bibitem[{Dener and Munson(2019)}]{Dener2019}
Dener A and Munson T (2019) \emph{Accelerating Limited-Memory Quasi-Newton Convergence for Large-Scale Optimization}.
\newblock Springer International Publishing.
\newblock ISBN 9783030227449, p. 495–507.
\newblock \doi{10.1007/978-3-030-22744-9_39}.

\bibitem[{{{E4S} {T}eam}(2024)}]{e4s:homepage}
{{E4S} {T}eam} (2024) {{E4S} {W}eb page}.
\newblock \url{https://e4s.io}.

\bibitem[{Faibussowitsch et~al.(2023)Faibussowitsch, Adams, Mills, Zampini and Zhang}]{faibussowitsch2023}
Faibussowitsch J, Adams MF, Mills RT, Zampini S and Zhang J (2023) Safe, seamless, and scalable integration of asynchronous {GPU} streams in {PETSc}.
\newblock \emph{arXiv preprint arXiv:2306.17801} .

\bibitem[{Falgout(2023)}]{hypre-users-manual}
Falgout R (2023) {hypre} users manual.
\newblock Technical Report Revision 2.28, Lawrence Livermore National Laboratory.
\newblock \urlprefix\url{https://hypre.readthedocs.io/}.

\bibitem[{Freund(1993)}]{freund1993tfqmr}
Freund RW (1993) A transpose-free quasi-minimal residual algorithm for non-hermitian linear systems.
\newblock \emph{SIAM Journal on Scientific Computing} 14(2): 470--482.
\newblock \doi{10.1137/0914029}.

\bibitem[{Hager et~al.(2016)Hager, Yoon, Ku, D{'}Azevedo, Worley and Chang}]{Hager2016}
Hager R, Yoon E, Ku SH, D{'}Azevedo EF, Worley PH and Chang CS (2016) A fully non-linear multi-species {F}okker{\textendash}{P}lanck{\textendash}{L}andau collision operator for simulation of fusion plasma.
\newblock \emph{Journal of Computational Physics} 315: 644--660.
\newblock \doi{10.1016/j.jcp.2016.03.064}.

\bibitem[{Hirvijoki and Adams(2017)}]{Hirvijoki2017}
Hirvijoki E and Adams MF (2017) Conservative discretization of the {Landau} collision integral.
\newblock \emph{Physics of Plasmas} 24(3): 032121.
\newblock \doi{10.1063/1.4979122}.

\bibitem[{Jolivet et~al.(2021)Jolivet, Roman and Zampini}]{jolivetromanzampini2020}
Jolivet P, Roman J and Zampini S (2021) {KSPHPDDM} and {PCHPDDM}: extending {PETSc} with advanced {Krylov} methods and robust multilevel overlapping {Schwarz} preconditioners.
\newblock \emph{Computers and Mathematics with Applications} 84: 277--295.

\bibitem[{{Khronos SYCL Working Group}(2020)}]{SYCL}
{Khronos SYCL Working Group} (2020) {SYCL} 2020 specification).
\newblock \urlprefix\url{https://registry.khronos.org/SYCL/specs/sycl-2020/pdf/sycl-2020.pdf}.

\bibitem[{Kothe et~al.(2019)Kothe, Lee and Qualters}]{ecp-kothe-lee-qualters-2019}
Kothe D, Lee S and Qualters I (2019) Exascale computing in the {United States}.
\newblock \emph{IEEE CiSE} 21(1): 17--29.
\newblock \doi{10.1109/MCSE.2018.2875366}.

\bibitem[{Liegeois et~al.(2023)Liegeois, Rajamanickam and Berger-Vergiat}]{Liegeois2023}
Liegeois K, Rajamanickam S and Berger-Vergiat L (2023) Performance portable batched sparse linear solvers.
\newblock \emph{IEEE Transactions on Parallel and Distributed Systems} 34(5): 1524--1535.
\newblock \doi{10.1109/TPDS.2023.3249110}.

\bibitem[{Liu and Nocedal(1989)}]{liu1989limited}
Liu DC and Nocedal J (1989) On the limited memory bfgs method for large scale optimization.
\newblock \emph{Mathematical programming} 45(1-3): 503--528.

\bibitem[{Mills et~al.(2021)Mills, Adams, Balay, Brown, Dener, Knepley, Kruger, Morgan, Munson, Rupp et~al.}]{mills2021toward}
Mills RT, Adams MF, Balay S, Brown J, Dener A, Knepley M, Kruger SE, Morgan H, Munson T, Rupp K et~al. (2021) Toward performance-portable {PETSc} for {GPU}-based exascale systems.
\newblock \emph{Parallel Computing} 108: 102831.

\bibitem[{{NVIDA}(2024)}]{CUDA}
{NVIDA} (2024) {CUDA C++} programming guide.
\newblock \urlprefix\url{https://docs.nvidia.com/cuda/pdf/CUDA_C_Programming_Guide.pdf}.

\bibitem[{NVIDIA(2024)}]{NVSHMEM}
NVIDIA (2024) {NVIDIA} {OpenSHMEM} library ({NVSHMEM}) documentation.
\newblock \urlprefix\url{https://docs.nvidia.com/nvshmem/api/index.html}.

\bibitem[{{Open Source Software Solutions, Inc.}(2020)}]{OpenSHMEM}
{Open Source Software Solutions, Inc} (2020) {OpenSHMEM} application programming interface v1.5.
\newblock \urlprefix\url{http://www.openshmem.org/}.

\bibitem[{{OpenMP Architecture Review Board}(2021)}]{OpenMP}
{OpenMP Architecture Review Board} (2021) {OpenMP} application programming interface.
\newblock \urlprefix\url{https://www.openmp.org/wp-content/uploads/OpenMP-API-Specification-5-2.pdf}.

\bibitem[{Panda et~al.(2024)}]{OSUMicro}
Panda D et~al. (2024) {OSU} microbenchmarks v7.3.
\newblock \emph{http://mvapich.cse.ohio-state.edu/benchmarks/} .

\bibitem[{Rupp et~al.(2016)Rupp, Tillet, Rudolf, Weinbub, Morhammer, Grasser, Jungel and Selberherr}]{VIENNACL}
Rupp K, Tillet P, Rudolf F, Weinbub J, Morhammer A, Grasser T, Jungel A and Selberherr S (2016) Vienna{CL}---linear algebra library for multi-and many-core architectures.
\newblock \emph{SIAM Journal on Scientific Computing} 38(5): S412--S439.

\bibitem[{Trott et~al.(2022)Trott, Lebrun-Grandié, Arndt, Ciesko, Dang, Ellingwood, Gayatri, Harvey, Hollman, Ibanez, Liber, Madsen, Miles, Poliakoff, Powell, Rajamanickam, Simberg, Sunderland, Turcksin and Wilke}]{KOKKOS}
Trott CR, Lebrun-Grandié D, Arndt D, Ciesko J, Dang V, Ellingwood N, Gayatri R, Harvey E, Hollman DS, Ibanez D, Liber N, Madsen J, Miles J, Poliakoff D, Powell A, Rajamanickam S, Simberg M, Sunderland D, Turcksin B and Wilke J (2022) Kokkos 3: Programming model extensions for the exascale era.
\newblock \emph{IEEE Transactions on Parallel and Distributed Systems} 33(4): 805--817.
\newblock \doi{10.1109/TPDS.2021.3097283}.

\bibitem[{Trotter et~al.(2023)Trotter, Langguth and Cai}]{trotter2023targeting}
Trotter JD, Langguth J and Cai X (2023) Targeting performance and user-friendliness: Gpu-accelerated finite element computation with automated code generation in fenics.
\newblock \emph{Parallel Computing} 118: 103051.

\bibitem[{{{xSDK} {T}eam}(2024)}]{xsdk:homepage}
{{xSDK} {T}eam} (2024) {{xSDK} {W}eb page}.
\newblock \url{https://xsdk.info}.

\bibitem[{Zampini et~al.(2024)Zampini, Zerbinati, Turkyyiah and Keyes}]{Zampini_PASC}
Zampini S, Zerbinati U, Turkyyiah G and Keyes D (2024) {PETScML}: Second-order solvers for training regression problems in scientific machine learning.
\newblock In: \emph{Proceedings of the Platform for Advanced Scientific Computing Conference}, PASC '24. New York, NY, USA: Association for Computing Machinery.
\newblock ISBN 9798400706394.
\newblock \doi{10.1145/3659914.3659931}.
\newblock \urlprefix\url{https://doi.org/10.1145/3659914.3659931}.

\bibitem[{Zhang et~al.(2022)Zhang, Constantinescu and Smith}]{Zhang2022tsadjoint}
Zhang H, Constantinescu EM and Smith BF (2022) {PETSc TSAdjoint:} a discrete adjoint {ODE} solver for first-order and second-order sensitivity analysis.
\newblock \emph{SIAM Journal on Scientific Computing} 44(1): C1--C24.
\newblock \doi{10.1137/21M140078X}.

\bibitem[{Zhang et~al.(2021)Zhang, Brown, Balay, Faibussowitsch, Knepley, Marin, Mills, Munson, Smith and Zampini}]{PetscSF_TPDS_2021}
Zhang J, Brown J, Balay S, Faibussowitsch J, Knepley M, Marin O, Mills RT, Munson T, Smith BF and Zampini S (2021) The {PetscSF} scalable communication layer.
\newblock \emph{IEEE Transactions on Parallel \& Distributed Systems} \doi{10.1109/TPDS.2021.3084070}.

\bibitem[{Zhang et~al.(2019)Zhang, Almgren, Beckner, Bell, Blaschke, Chan, Day, Friesen, Gott, Graves, Katz, Myers, Nguyen, Nonaka, Rosso, Williams and Zingale}]{AMReX_JOSS}
Zhang W, Almgren A, Beckner V, Bell J, Blaschke J, Chan C, Day M, Friesen B, Gott K, Graves D, Katz M, Myers A, Nguyen T, Nonaka A, Rosso M, Williams S and Zingale M (2019) {AMReX}: a framework for block-structured adaptive mesh refinement.
\newblock \emph{Journal of Open Source Software} 4(37): 1370.
\newblock \doi{10.21105/joss.01370}.

\bibitem[{Zhou et~al.(2022)Zhou, Raffenetti, Guo and Thakur}]{zhou2022mpix}
Zhou H, Raffenetti K, Guo Y and Thakur R (2022) Mpix stream: An explicit solution to hybrid mpi+ x programming.
\newblock In: \emph{Proceedings of the 29th European MPI Users' Group Meeting}. pp. 1--10.

\end{thebibliography}

\begin{section}{Author biographies}
Richard Tran Mills is a Computational Scientist in the Mathematics and Computer Science Division at Argonne National Laboratory.
His research spans high-performance scientific computing, geospatiotemporal data mining and machine learning, computational hydrology, and climate change science.
He is one of the original developers of PFLOTRAN, an open-source code for massively parallel simulation of hydrologic flow and reactive transport problems, and is a core developer of PETSc, the Portable, Extensible Toolkit for Scientific Computation.
He earned his Ph.D. in Computer Science in 2004 at the College of William and Mary, where he was a Department of Energy Computational Science Graduate Fellow. Prior to that, he studied geology and physics at the University of Tennessee, Knoxville as a Chancellor's Scholar.

Mark Adams is a Staff Scientist in the Scalable Solvers Group at Lawrence Berkeley National Laboratory (LBNL), where he develops the algebraic multigrid solver in PETSc and structure preservation techniques for kinetic equations. He holds a Ph.D. in Civil Engineering from U.C. Berkeley.

Satish Balay is a software engineer at Argonne National Laboratory. He received his M.S. in computer science from Old Dominion University. He is a developer of PETSc.

Jed Brown is an associate professor of computer science at the University of Colorado Boulder. He received his Dr.Sc. from ETH Z\"urich and BS+MS from the University of Alaska Fairbanks. He is a maintainer of PETSc and leads a research group on fast algorithms and community software for physical prediction, inference, and design.

Jacob Faibussowitsch is a Senior Software Engineer at NVIDIA. Previously, he was a Research Software Engineer at Argonne National Laboratory, working as a full-time developer of PETSc. He received his M.S. in mechanical engineering at the University of Illinois at Urbana-Champaign, where is he also received his B.S. His work focuses on high-performance GPU software development.

Toby Isaac is computational mathematician at Argonne National Laboratory.  He received his Ph.D. in computational science, mathematics, and engineering from the University of Texas at Austin.  He is a developer of PETSc.

Matthew Knepley is a professor in the University at Buffalo. He received his Ph.D. in computer science from Purdue University and his B.S. from Case Western Reserve University. His work focuses on computational science, particularly in geodynamics, plasma physics, subsurface flow, and molecular mechanics. He is a maintainer of PETSc.

Todd Munson is a senior computational scientist at Argonne National Laboratory and the Software Ecosystem and Delivery Control
Account Manager for the U.S.\ DOE Exascale Computing Project. His interests range from numerical methods for nonlinear optimization and variational 
inequalities to workflow optimization for online data analysis and reduction. He is a developer of the Toolkit for Advanced Optimization.

Hansol Suh is a Predoctoral Appointee in the Mathematics and Computer Science Division at Argonne National Laboratory. He is a developer of PETSc. He earned his M.S. in Computational Science and Engineering in 2022 at the Georgia Institute of Technology.

Stefano Zampini is a senior research scientist in the Computer, Electrical and Mathematical Sciences and Engineering division of King Abdullah University for Science and Technology (KAUST), Saudi Arabia. He received his Ph.D. in applied mathematics from the University of Milano Statale. He is a developer of PETSc. His contributions to the open-source software community for CSE extend to other frameworks in the US DOE ecosystem, namely the MFEM and deal.II libraries for finite-element-based simulations. He is also a member of the HPC technical committee of the CFD software package OpenFOAM.

Hong Zhang is an applied mathematics specialist at Argonne National Laboratory. He received his Ph.D. in computer science from Virginia Tech. His research focuses on scientific machine learning and high performance computing. He is a developer of PETSc.

Junchao Zhang is a software engineer at Argonne National Laboratory. He received his Ph.D. in computer science from the Chinese Academy of Sciences, Beijing, China. He is a developer of PETSc.
\end{section}

\end{document}